\documentclass[journal,draftcls,onecolumn,letterpaper,11pt]{IEEEtran}
\usepackage{cite,amsmath,amssymb}
\usepackage{citesort}
\usepackage{graphicx}
\usepackage{algorithm,algorithmic}
\usepackage{subfigure}
\usepackage{mathrsfs}
\usepackage{url}
\usepackage{bbm}
\usepackage{amsthm}
\usepackage{setspace}
\usepackage{color}
\newtheorem{theorem}{Theorem}

\newtheorem{definition}{Definition}

\newtheorem{proposition}{Proposition}

\newtheorem{lemma}{Lemma}

\newtheorem{remark}{Remark}

\makeatletter
\newcounter{parentalgorithm}

\newcommand{\Prob}{\mathbb{P}}

\newcommand{\bla}{{\boldsymbol\lambda}}

\newcommand{\la}{\lambda}

\newcommand{\bs}{{\bf s}}

\newcommand{\bnu}{{\boldsymbol\nu}}

\newcommand{\E}{\mathbb{E}}
\newcommand{\T}{\mathsf{T}}
\newcommand{\MC}{\mathcal{C}}
\newcommand{\MR}{\mathcal{R}}
\newcommand{\MG}{\mathcal{G}}
\newcommand{\MD}{\mathcal{D}}
\newcommand{\ML}{\mathcal{L}}

\newcommand{\MH}{\mathcal{H}}
\newcommand{\MV}{\mathcal{V}}
\newcommand{\MA}{\mathcal{A}}

\newcommand{\ignore}[1]{{}}
\newcommand{\Bdelta}{{\boldsymbol\delta}}
\newcommand{\bmu}{{\boldsymbol\mu}}
\newcommand{\lb}{\left(}
\newcommand{\rb}{\right)}
\begin{document}
\title{Sequential Hypothesis Test with Online Usage-Constrained Sensor Selection}
\author{Shang Li$^*$, Xiaoou Li$^\dag$, Xiaodong Wang$^*$, Jingchen Liu$^\dag$
\thanks{$^*$S. Li and X. Wang are with Department of Electrical Engineering, Columbia University, New York, NY 10027 (e-mail: \{shang,wangx\}@ee.columbia.edu).
}
\thanks{$^\dag$X. Li and J. Liu are with Department of Statistics, Columbia University, New York, NY 10027 (e-mail: \{xiaoou,jcliu\}@stat.columbia.edu).
}}
\maketitle

\begin{abstract}

This work investigates the sequential hypothesis testing problem with online sensor selection and sensor usage constraints. That is, in a sensor network, the fusion center sequentially acquires samples by selecting one ``most informative'' sensor at each time until a reliable decision can be made. In particular, the sensor selection is carried out in the online fashion since it depends on all the previous samples at each time. Our goal is to develop the sequential test (i.e., stopping rule and decision function) and sensor selection strategy that minimize the expected sample size subject to the constraints on the error probabilities and sensor usages. To this end, we first recast the usage-constrained formulation into a Bayesian optimal stopping problem with different sampling costs for the usage-contrained sensors. The Bayesian problem is then studied under both finite- and infinite-horizon setups, based on which, the optimal solution to the original usage-constrained problem can be readily established. Moreover, by capitalizing on the structures of the optimal solution, a lower bound is obtained for the optimal expected sample size. In addition, we also propose algorithms to approximately evaluate the parameters in the optimal sequential test so that the sensor usage and error  probability constraints are satisfied.  Finally, numerical experiments are provided to illustrate the theoretical findings, and compare with the existing methods.



\end{abstract}
\begin{IEEEkeywords}
Sequential hypothesis test, online sensor selection, reliability, sensor usages, dynamic programming.
\end{IEEEkeywords}

\newpage
\section{Introduction}

Nowadays the sequential hypothesis test has been widely applied in many   applications because it generally requires smaller sample size on average compared to its fixed-sample-size counterpart. Notably, \cite{Wald48} proved that the sequential probability ratio test (SPRT) yields the minimum expected sample size under both null and alternative hypotheses given the error probabilities. Since this pioneering work, a rich body of studies on the sequential test have emerged under different circumstances \cite{SeqA_book}. One of the most important applications of sequential test is found in sensor networks \cite{Veeravalli93,Tsitsiklis93,Blum97,Mei08,LiLi15}.  In this work, we consider the sequential hypothesis test when sensor access at the fusion center is restricted, and efficient sensor scheduling/selection is of interest.  That is, the sensor network with different types of sensors (i.e., heterogenous sensors) and a fusion center aims to test between two hypotheses; however, only one of the available sensors can take samples  and communicate with the fusion center at each sampling instant. Such a setup often arises when the fusion center possessses limited processing capability/resources, or the sensors contradict/exclude one another. For instance, the echo-based sensors like sonar sensors can interfere with each other \cite{Gupta06}. In practice, the heterogenous sensors could also refer to multiple information resources, and the processing unit (i.e., fusion center) can only analyze one at a time. This model well describes, for example, the human decision process. As such, in order to reach a quick and reliable decision, strategically selecting the ``most informative'' sensor, which often depends on the parameter values or the true hypothesis that is unknown, has become the pivotal problem.   


In the context of fixed-sample-size statistical inference, sensor selection has been well studied, mainly from the optimization standpoint.  In particular, \cite{Gupta06} proposed a random selection scheme to minimize the error covariance of a process tracking problem; for the Kalman filter, \cite{Mo11} devised a multi-stage strategy to select a subset of sensors so that an objective function related to the  error covariance matrix was minimized; \cite{Joshi09} put forth a convex-optimization-based approach to select multiple sensors for the parameter estimation in linear system. For the fixed-sample-size hypothesis test, \cite{Bajovic11} investigated sensor scheduling based on information-metric criteria such as Kullback-Leibler and Chernoff distances.  

The studies on the sensor selection for sequential hypothesis test have mainly branched into the offline (a.k.a. open-loop) and online (a.k.a. closed-loop) approaches. The former category essentially involves independent random selection over time, with the probability preassigned to each sensor. Along this direction,   \cite{Srivastava11,Srivastava11_2} introduced random sensor selection to the multi-hypothesis sequential probability ratio test (MSPRT), and designed the selection probability such that its approximate decision delay was minimized. They concluded that the optimal random selection strategy involve at most two sensors for binary-hypothesis test. Namely, the fusion center should either always use one sensor, or randomly switch between two sensors, and disregard the rest. Similar teniques were later applied to the quickest detection with stochastic surveillance control \cite{Srivastava13}. Recently, focusing on the binary-hypothesis test, \cite{Bai15} further imposed constraints on the sensor usages, i.e., sensors, on average, cannot be selected more than their prescribed limits, and obtained the selection probabilities for SPRT with random sensor selection. 


Despite their simple implementations, the open-loop approaches do not make use of the accumulating sample information, thus are suboptimal in general. On the contrary, the online approaches take all previous samples into account at each step for sensor selection, and generally yield superior performance. As a matter of fact, dynamic sensing control is one of the major advantages of sequential processing. To this end, \cite{Chernoff59} selected the sensor that was most informative under the most likely true hypothesis at each step. \cite{Javidi10,Javidi13,Javidi13_2} investigated the sequential multi-hypothesis test with observation control, and provided lower and upper bound for its asymptotic performance. Two asymptotically optimal  algorithms were proposed there. The variant of sequential hypothesis test---changepoint detection with observation control were considered by \cite{Banerjee12,Banerjee13} based on Bayesian and non-Bayesian settings respectively. Meanwhile, \cite{Kumar08} assumed identical sensors, and studied the Bayesian changepoint detection with control on the number of active sensors. Most of the above online approaches are based on heuristics and perform well in the asymptotic regime, where error probabilities are extremely low. 
On the other hand, focusing on the non-asymptotic regime, \cite{Bai14} considered the online sensor selection strategy for the SPRT. However, it aimed to minimize the decision delay given that SPRT was used. 
Instead, the recent work \cite{Xiaoou15} jointly solved a Bayesian hypothesis testing problem for both the optimal sequential test and online selection strategy. 

In this work, we also aim for the optimal sequential test and online sensor selection simultaneously. Moreover, we further introduce the constraints on the sensor usages into the formulation, which would potentially embrace a much wider range of practical problems. That is, certain sensors in the network are not allowed to be selected more than a prescribed number of times on average. The usage constraints naturally arise when one intends to restrain the sensors from being overused due to their limited battery/lifetime, or if the fairness for all sensors in the network is important \cite{Bai15}. We summarize the contributions as follows:
\begin{itemize}
\item To the best of our  knowledge, this is the first work that jointly solves for the optimal sequential test and online sensor selection when sensor usage constraints are considered. Moreover, this work distinguishes from \cite{Bai15}, where the usage-contrained sensor selection is also studied, in terms of its online/closed-loop setup.


\item Note that most of the existing works on sensor selection for sequential test only apply to infinite-horizon, where sample size (or decision delay) at a specific realization can go to infinity if necessary. This may not be realistic in some applications. In contrast, we consider {\it both} the infinite-horizon and finite-horizon scenarios. In the later case, a fixed upper bound is imposed on the random sample size at every realizations.

\item We propose practical algorithm to systematically evaluate the parameters in the optimal sequential test and selection strategy. As long as the test performance constraints and the sensor usage constraints remain the same, this algorithm only needs to be run once offline. That is, once the parameters are calculated, they can be stored at the fusion center, based on which, the sequential test can be easily implemented.
\end{itemize}

The reminder of the paper is organized as follows. We first formulate the usage-constrained sequential hypothesis test in Section II. Then the optimal sequential test and sensor selection strategy are derived in Section III. In Section IV, we propose practical algorithms to design the parameters in the optimal scheme. Section V provides  numerical results to illustrate the theoretical results, and to compare with the offline random selection scheme. Finally, Section VI concludes this paper.
\section{Problem Formulation}

Consider a system consisting of $K$ sensors and a fusion center that aims to test between two hypotheses, whose priors are given as $\Prob\lb\mathcal{H}=i\rb=\pi_i, \, i=0, 1$. At each time instant, the fusion center selects one sensor to take a sample that is sent to the fusion center. This process continues until a reliable decision can be made. It is assumed that the fusion center possesses the statistical characterization of all sensors. That is, the conditional probability density functions $f_{\mathcal{H}}^\ell(x)$ of the random samples collected by sensor $\ell, \,\ell=1, 2, \ldots, K$ are known to the fusion center. Without loss of generality, we assume that the sensor network is heterogenous, i.e., there are no two sensors with identical $f_{\mathcal{H}}^\ell(x)$'s. In addition, the random samples are assumed to be independent and identically distributed (i.i.d.) over time for the same sensor $\ell$, and independent across different sensors. 

On one hand, if there is a dominant sensor that always outperforms all other sensors, the fusion center should always use it in the absence of usage constraint.  Then the problem reduces to a single-sensor sequential hypothesis test, and the SPRT yields the quickest decision. One such example is the test between zero ($\MH_0$) and non-zero Gaussian means ($\MH_1$), where the sensor with the largest mean shift under $\MH_1$ should prevail. On the other hand, the efficiency of a sensor generally depends on the true hypothesis. For example, some  sensors can be more informative under $\MH_0$ and less so under $\MH_1$, thus accelerating the decision speed when $\mathcal{H}_0$ is true, and slowing down the decision speed otherwise.  
Moreover, even the dominant sensor cannot be used all the time if its usage is restrained. 
In general, the online sensor selection procedure is performed  based on the  accumulated  sample information, which is explained as follows.

There are three essential operations in the online procedure:
\begin{enumerate}
\item Sensor selection strategy: Let $\Pi\triangleq\{1, 2, \ldots, K\}$ be the set of all sensors, and $\{X_1, \ldots, X_t\}$ denote the sequence of samples received at the fusion center. Then the sensor selected at time $t$ can be defined as $\delta_t: \{X_1, \ldots, X_{t-1}\}\to j\in \Pi$.  In addition, we denote the sequence of sensor selections from time $i$  to time $j$ as $\Bdelta_{i:j}$, and  $\Bdelta_{i:j}\triangleq \emptyset$ if $i>j$.  Note that since at any time,  the distribution of the next sample depends on the selection function, the fusion center observes dependent random samples $\{X_t\}$.
\item Stopping rule: The random sample size is characterized by the stopping time  $\T$. In specific, the event $\{\T=t\}$ means that the sample size is equal to $t$, which depends on $\{X_1, \ldots, X_t\}$. In this work, we focus on the deterministic stopping rule, i.e., $\Prob\lb\T=t|X_1, \ldots, X_t\rb$ is either zero or one.
\item Decision function: Upon stopping at $\T=t$, a final decision between the two hypotheses is made, $D_t: \{X_1, \ldots, X_t\}\to \{0, 1\}$.
\end{enumerate}
As such, the fusion center is faced with the following hypothesis testing problem:
\begin{align*}
\begin{array}{ll}
\mathcal{H}_0: &X_t\sim f^{\delta_t}_0(x), \quad t=1, 2, \ldots \\
\mathcal{H}_1: & X_t\sim f^{\delta_t}_1(x), \quad t=1, 2, \ldots.
\end{array}
\end{align*}
The performance indicators for sequential hypothesis test include the expected sample size and the error probabilities. In particular, the expected sample size $\E\T=\pi_0\E_0\lb\T\rb+\pi_1\E_1\lb\T\rb$ is the weighted sum of the conditional expected sample sizes, and the type-I and type-II error probabilities are $\Prob_0\lb D_\T=1\rb$ and $\Prob_1\lb D_\T=0\rb$ respectively\footnote{One can also use the weighted sum of type-I and type-II error rates as the error probability. Here we adopt the formulation in \cite{Bai15}, and consider them individually. Nevertheless, the method developed in this work can be applied to the former case.}. 
Here the expectation $\E\lb\cdot\rb$ is taken over the joint distribution of $\mathcal{H}$ and $X_t$, and $\E_i\lb\cdot\rb$ is taken over the  distribution of $X_t$ conditioned on $\{\mathcal{H}=i\}$. 

Moreover, we also impose constraints on the usage of sensors. Denote $\Omega$ as the set of sensors whose usages are restrained. Then for each sensor $\ell\in \Omega$, the average number of times that sensor $\ell$ is selected, $\E\lb \sum_{t=1}^\T \mathbbm{1}_{\{\delta_t=\ell\}}\rb$, is constrained to be no greater than $T^\ell\in \mathbb{R}^+$.  As such, we arrive at the following constrained sequential problem:
\begin{align}\label{P1}
\begin{array}{ll}
\min_{\{\Bdelta_{1:\T}, D_\T, \T\}} & \E \T \\
\text{subject to} & \ignore{\E\lb\mathbbm{1}_{\{D_\T\neq \MH\}}\rb}\Prob_0\lb D_\T=1\rb\le \alpha,\, \Prob_1\lb D_\T=0\rb\le \beta,\\ \phantom{\text{subject to}} & \E\lb \sum_{t=1}^\T \mathbbm{1}_{\{\delta_t=\ell\}}\rb\le T^\ell,  \quad \ell\in \Omega.
\end{array}\tag{P1}
\end{align}
In the following sections, we will solve \eqref{P1} under both the finite-horizon and infinite-horizon setups. The finite-horizon setup imposes an upper bound on $\T$ for any realization, beyond which no sample can be taken; whereas the infinite-horizon setup allows the sequential test to continue as long as the termination condition is not met. In addition to its relevance in many applications, the finite-horizon case can also be used as a building block for the infinite-horizon problem. For notational convenience, we define the class of infinite-horizon procedures:
\begin{align}
{\bf C}\lb \alpha, \beta, \{T^\ell\}_{\ell\in \Omega}\rb\triangleq &\Big\{\{\Bdelta_{1:\T}, D_\T, \T\}: \, \Prob_0\lb D_\T=1\rb\le \alpha,\, \nonumber\\& \Prob_1\lb D_\T=0\rb\le \beta, \;\text{and}\;\; \E\lb \sum_{t=1}^\T \mathbbm{1}_{\{\delta_t=\ell\}}\rb\le T^\ell,  \ell\in \Omega\Big\},
\end{align}
and the class of finite-horizon procedures:
\begin{align}
{\bf C}_N\lb \alpha, \beta, \{T^\ell\}_{\ell\in \Omega}\rb\triangleq \Big\{\{\Bdelta_{1:\T}, D_\T, \T\}\in {\bf C}\lb \alpha, \beta, \{T^\ell\}_{\ell\in \Omega}\rb: \T\le N\Big\}.
\end{align}
Our goal is to find the optimal triplets $\{\Bdelta_{1:\T}, \T, D_\T\}$ that yield the smallest expected sample sizes $\E\T$ in the classes ${\bf C}_N\lb \alpha, \beta, \{T^\ell\}_{\ell\in \Omega}\rb$ and ${\bf C}\lb \alpha, \beta, \{T^\ell\}_{\ell\in \Omega}\rb$ respectively. 

\section{Optimal Sequential Test with Constrained Online Sensor Selection}
In this section, we first recast \eqref{P1} into an unconstrained optimal stopping problem, which we then solve under both finite-horizon and infinite-horizon setups. The solutions lead us to the optimal sequential solutions to the original constrained problem \eqref{P1}. 

By introducing Lagrange multipliers to \eqref{P1}, we arrive at the following Bayes objective function:
\begin{align}\label{Bayes_obj}
\MR(\Bdelta_{1:\T}, D_\T, \T)&\triangleq \E\T+\mu_0\pi_0\Prob_0\lb{D_\T=1}\rb+\mu_1\pi_1\Prob_1\lb{D_\T=0}\rb+\sum_{\ell\in\Omega}\lambda_\ell \,\E\lb \sum_{t=1}^\T\mathbbm{1}_{\{\delta_t=\ell\}}\rb\nonumber\\&=\E\lb \T+\mu_0\mathbbm{1}_{\{D_\T=1; \MH=0\}}+\mu_1\mathbbm{1}_{\{D_\T=0; \MH=1\}}+\sum_{\ell\in\Omega}\lambda_\ell \lb \sum_{t=1}^\T\mathbbm{1}_{\{\delta_t=\ell\}}\rb\rb\nonumber\\&={\E\lb \sum_{t=1}^\T \underbrace{\lb 1+\mathbbm{1}_{\{\delta_t\in \Omega\}}\lambda_{\delta_t}\rb}_{\MC_{\delta_t}}+\underbrace{\mu_0\mathbbm{1}_{\{D_\T=1; \MH=0\}}+\mu_1\mathbbm{1}_{\{D_\T=0; \MH=1\}}}_{\mu\lb D_\T, \MH\rb}\rb}.
\end{align}
Note that $\MC_j\triangleq 1+\lambda_j$ and $\lambda_j\ge 0$ for $j\in \Omega$, and $\MC_j\triangleq 1$ for $j\notin \Omega$. 


\subsection{Finite-Horizon Solution to the Bayes Problem}
In this subsection, under the finite-horizon setup, we aim to find the optimal sensor selection, stopping time and decision rule such that the Bayes risk in \eqref{Bayes_obj} is minimized, i.e., 
\begin{align}\label{P1_Bayes}
\min_{\{\Bdelta_{1:\T}, D_\T, \T\}, \T\le N}\quad \mathcal{R}\lb\Bdelta_{1:\T}, D_\T, \T\rb=\E\lb\sum_{t=1}^\T\MC_{\delta_t}+\mu\lb D_\T, \MH\rb\rb.
\end{align}
Define the cumulative log-likelihood ratio (LLR)
\begin{align}
L_n\triangleq \sum_{t=1}^n\underbrace{\log \frac{f_1^{\delta_t}(X_t)}{f_0^{\delta_t}(X_t)}}_{l_{\delta_t}(X_t)},
\end{align}
and the posterior probabilities $\pi_i(t)\triangleq\Prob\lb\MH=i|X_{1:t}, \Bdelta_{1:t}\rb$, $i\in\{0, 1\}$ with $\pi_i(0)=\pi_i$. These two statistics relate to each other as follows
\begin{align}\label{posterior}
\pi_1(n)=\frac{\pi_1e^{L_n}}{\pi_0+\pi_1e^{L_n}}=\frac{\pi_1({n-1})e^{l_{\delta_n}}}{\pi_0({n-1})+\pi_1({n-1})e^{l_{\delta_n}}},\quad  L_n=\log\frac{\pi_0\pi_1(n)}{\pi_1\pi_0(n)}.
\end{align}
\subsubsection{Decision Function} We begin with solving the terminal decision function. Since
\begin{align}
\mathcal{R}\lb\Bdelta_{1:\T}, D_\T, \T\rb =&\E\lb \sum_{t=1}^\T{\MC_{\delta_t}}\rb+\sum_{t=1}^\infty \E\left[\mathbbm{1}_{\{\T= t\}}\lb\mu_0\mathbbm{1}_{\{D_\T=1; \MH=0\}}+\mu_1\mathbbm{1}_{\{D_\T=0; \MH=1\}}\rb\right]\nonumber\\ =&\E\lb \sum_{t=1}^\T{\MC_{\delta_t}}\rb+\sum_{t=1}^\infty \E\lb\E_\MH\lb\left.\mu_0\mathbbm{1}_{\{D_t=1; \MH=0\}}+\mu_1\mathbbm{1}_{\{D_t=0; \MH=1\}}\right|X_{1:t},\Bdelta_{1:t}\rb\mathbbm{1}_{\{\T=t\}}\rb\nonumber\\ =&\E\lb \sum_{t=1}^\T{\MC_{\delta_t}}\rb+\sum_{t=1}^\infty \E\left[\lb\mu_0\pi_0(t)\mathbbm{1}_{\{D_t\neq 0\}}+\mu_1\pi_1(t)\mathbbm{1}_{\{D_t\neq 1\}}\rb\mathbbm{1}_{\{\T=t\}}\right],\label{opt_decision}
\end{align}
we have $D_t^\star=\mathbbm{1}_{\{\mu_0\pi_0(t)\le \mu_1\pi_1(t)\}}$ given $\T=t$, i.e., 
\begin{align}\label{DecRule}
D_\T^\star=\mathbbm{1}_{\{\mu_0\pi_0(\T)\le \mu_1\pi_1(\T)\}}.
\end{align}
\subsubsection{Selection Strategy and Stopping Rule} For notational convenience, define the class 
\begin{align}\label{classA}
\MA_n^N\triangleq \left\{\{\Bdelta_{n+1:\T}, \T\}: n\le \T\le N\right\},
\end{align} 
in which the procedures do not stop before $n$ and can not go beyond $N$. By substituting $D_\T$ with \eqref{DecRule}, \eqref{P1_Bayes} becomes
\begin{align}\label{Opt_stopping}
\min_{\{\Bdelta_{1:\T}, \T\}\in \MA^N_0}\;  \E\lb \sum_{t=1}^\T \MC_{\delta_t} + \underbrace{\min\left\{\mu_0\pi_0(\T), \mu_1\pi_1(\T)\right\}}_{\phi(\pi_1(\T))}\rb,
\end{align}
where $\phi(x)\triangleq \min\{\mu_1x, \mu_0(1-x)\}$. We next solve \eqref{Opt_stopping} to obtain the optimal  sensor selection strategy and stopping rule.

Define the optimal cost of the procedures that do not stop before $t=n$, i.e.,  the  ``cost-to-go'' function
\begin{align}\label{V_n_N}
\MV^N_{n}\lb X_{1:n}, \Bdelta_{1:n}\rb&\triangleq\min_{\{\Bdelta_{n+1:\T}, \T\}\in \MA^N_n} \E\lb\left. \sum_{t=1}^\T\MC_{\delta_t}+\phi\lb\pi_1(\T)\rb\right|X_{1:n},\Bdelta_{1:n}\rb
\end{align}
Note that $\MV_0^N$ (which is not a function of any samples)  is equal to \eqref{Opt_stopping} by definition and $\MV_N^N(X_{1:N},\Bdelta_{1:N})=\phi\lb \pi_1(N)\rb+\sum_{t=1}^N \MC_{\delta_t}$ since the test has to stop at $N$ if not before it.  Invoking the technique of dynamic programming, the cost-to-go \eqref{V_n_N} can be recursively solved by the following backward recursion \cite{SeqA_book}:
\begin{align}\label{Backward}
&{\MV}_n^N(X_{1:n}, \Bdelta_{1:n})=
\min\left\{\underbrace{\phi\lb \pi_1(n)\rb+\sum_{t=1}^n\MC_{\delta_t}}_{r_s\lb X_{1:n},\Bdelta_{1:n}\rb},\; \underbrace{\min_{\delta_{n+1}}\left[\E\lb\left.\MV_{n+1}^N\lb X_{1:n+1}, \Bdelta_{1:n+1}\rb\right|X_{1:n}, \Bdelta_{1:n}\rb\right]}_{r_c\lb X_{1:n},\Bdelta_{1:n}\rb}\right\},
\end{align}
with $n=N-1, N-2, \ldots, 1, 0$. According to the principle of optimality, the optimal stopping time happens when the cost of stopping at the present instant is lower than the expected cost of continuing \cite{Wald48,Ferguson_book}, i.e., $\T^\star=\min\{n: g_n(X_{1:n}, \Bdelta_{1:n})\triangleq r_s\lb X_{1:n},\Bdelta_{1:n}\rb-r_c\lb X_{1:n},\Bdelta_{1:n}\rb \le 0\}$,  where
\begin{align}\label{stopping_rule}
g_n&\lb X_{1:n}, \Bdelta_{1:n}\rb= \phi\lb \pi_1(n)\rb+\sum_{t=1}^n\MC_{\delta_t}-\min_{\delta_{n+1}}\left[\E\lb\left.\MV_{n+1}^N\lb X_{1:n+1}, \Bdelta_{1:n+1}\rb\right|X_{1:n}, \Bdelta_{1:n}\rb\right]\nonumber\\&=\phi\lb \pi_1(n)\rb-\min_{\delta_{n+1}} \left\{\MC_{\delta_{n+1}}+
\min_{\{\Bdelta_{n+2:\T}, \T\}\in \MA^N_{n+1}}\left[\E\lb \left.\phi(\pi_1(\T))+\sum_{t=n+2}^\T \MC_{\delta_t}\right|X_{1:n}, \Bdelta_{1:n}\rb\right]\right\},
\end{align}
where the second equality is due to the definition of $\MV_n^N$ in \eqref{V_n_N}.

In theory, \eqref{Backward} and $\T^\star$ fully characterize the optimal stopping rule and selection strategy from the first to the $N$-th steps. However, this result is of limited practical value due to the high complexity brought by the high-dimensional quantities (i.e., $X_{1:n}$ and $\Bdelta_{1:n}$).  To this end, the following lemma significantly simplifies $\T^\star$ and \eqref{stopping_rule}, since it states that the hypothesis posterior (or equivalently, the LLR) is the sufficient statistic for the optimal stopping rule.  
\begin{lemma}\label{lemma:1}
The optimal stopping rule for \eqref{P1_Bayes} is a function of time and   hypothesis posterior, i.e., a time-variant function of the posterior, $\T^\star=\min\{n: g_n(\pi_1(n))\le0\}$.
\end{lemma}

\proof
See Appendix.
\endproof
The important implication of Lemma \ref{lemma:1} is that the selection strategy, which depends on all previous samples, can be summarized into a more compact form.  
\begin{lemma}
The optimal selection strategy for \eqref{P1_Bayes} is characterized by a time-variant function of the hypothesis posterior (or equivalently, the LLR), i.e.,  $\delta_{n+1}^\star=\psi_{n+1}(\pi_1(n))$. 
\end{lemma}
\proof
From \eqref{Backward}, the optimal selection strategy for $t=n+1$ is
\begin{align}\label{Lemma2_1}
\delta_{n+1}^\star&=\arg\,\min_{\delta_{n+1}}\, \E\lb\left.\MV_{n+1}^N\lb X_{1:n+1}, \Bdelta_{1:n+1}\rb\right|X_{1:n}, \Bdelta_{1:n}\rb,
\end{align}
and, by its definition, we have
\begin{align}\label{Lemma2_2}
\mathcal{V}_{n+1}^N\lb X_{1:n+1},\Bdelta_{1:n+1}\rb&=\min \left\{r_s\lb X_{1:n+1},\Bdelta_{1:n+1}\rb, r_c\lb X_{1:n+1},\Bdelta_{1:n+1}\rb\right\}
\nonumber\\&=\min\left\{0, -g_{n+1}(\pi_1(n+1))\right\}+r_s\lb X_{1:n+1},\Bdelta_{1:n+1}\rb\nonumber\\&=\phi\lb\pi_1(n+1)\rb+\sum_{t=1}^{n+1}\mathcal{C}_{\delta_t}-\max\left\{g_{n+1}(\pi_1(n+1)),0\right\}. 
\end{align}
Substituting \eqref{Lemma2_2} into \eqref{Lemma2_1} and neglecting the term $\sum_{t=1}^n\mathcal{C}_{\delta_t}$ that is independent of $\delta_{n+1}$, we arrive at
\begin{align}
\delta_{n+1}^\star&=\text{arg}\,\min_{\delta_{n+1}}\left\{\MC_{\delta_{n+1}}+ {\E \left.\Big[ \phi\lb \pi_1(n+1)\rb-\max\left\{g_{n+1}\lb \pi_1(n+1)\rb, 0\right\}\right|X_{1:n},\Bdelta_{1:n}\Big]} \right\}\nonumber\\&=\text{arg}\,\min_{\delta_{n+1}}\left\{\MC_{\delta_{n+1}}+ \underbrace{\E \left.\Big[ \phi\lb \pi_1(n+1)\rb-\max\left\{g_{n+1}\lb \pi_1(n+1), n+1\rb, 0\right\}\right|\pi_1(n)\Big]}_{u_n\lb \pi_1(n), \delta_{n+1}\rb} \right\}.
\end{align}
Note that the fact that the expectation term in the bracket is a time-variant function of $\pi_1(n)$ and $\delta_{n+1}$ (i.e., $u_n\lb \pi_1(n), \delta_{n+1}\rb$) follows from the relation between $\pi_1(n)$ and $\pi_1(n+1)$ given by \eqref{posterior}. Then $\delta_{n+1}^\star=\arg\min_{\delta} \;\widetilde u_n\lb\pi_1{(n)},\delta\rb\triangleq \mathcal{C}_{\delta}+u_n\lb \pi_1(n),\delta\rb$ which implies that the optimal selection is a time-variant function of the posterior, i.e., $\delta_{n+1}^\star=\psi_{n+1}\lb \pi_1(n)\rb$.  
\endproof
This result agrees with the intuition. Since the sensor efficiency depends on the actual hypothesis, it is reasonable to base the sensor selection upon the present belief (i.e., posterior) on the hypothesis.

Next we continue to study the stopping rule $\T^\star$ in more details. 
Define
\begin{align}
\MG^N_n(X_{1:n}, \Bdelta_{1:n})&\triangleq\MV_n^N\lb X_{1:n}, \Bdelta_{1:n}\rb-\sum_{t=1}^n\MC_{\delta_t}\nonumber\\&=\min_{\{\Bdelta_{n+1:\T}, \T\}\in \MA_n^N} \E\lb\left. \sum_{t=n+1}^\T\MC_{\delta_t}+\phi\lb\pi_1(\T)\rb\right|X_{1:n}, \Bdelta_{1:n}\rb.\label{def_G}
\end{align}
Meanwhile, $\MG_n^N(X_{1:n}, \Bdelta_{1:n})$ can be written as a function of $\pi_1(n)$ by using \eqref{Lemma2_2} as
\begin{align}\label{G_pi}
\MG_n^N(X_{1:n}, \Bdelta_{1:n})
&=\phi\lb\pi_{1}(n)\rb-\max\left\{g_n(\pi_1(n)),0\right\}=\MG^N_n\lb\pi_1(n)\rb,
\end{align}
where $\MG_N^N(X_{1:N}, \Bdelta_{1:N})=\phi\lb\pi_1(N)\rb$. 


Then, by substracting $\sum_{t=1}^n\mathcal{C}_{\delta_t}$ on both sides of \eqref{Backward}, we obtain
\begin{align}
r_s-\sum_{t=1}^n\mathcal{C}_{\delta_t}&=\phi(\pi_1(n)),\label{G_0}\\
\text{and}\quad r_c\lb X_{1:n},\Bdelta_{1:n}\rb-\sum_{t=1}^n\mathcal{C}_{\delta_t}&=\min_{\delta_{n+1}}\;\E\left[\left.\MV_{n+1}^N\lb X_{1:n+1}, \Bdelta_{1:n+1}\rb-\sum_{t=1}^n\mathcal{C}_{\delta_t}\right|X_{1:n},\Bdelta_{1:n}\right]\nonumber\\&=\min_{\delta_{n+1}}\;\E\left[\left.\mathcal{C}_{\delta_{n+1}}+\MG^N_{n+1}\lb \pi_1(n+1)\rb\right|X_{1:n},\Bdelta_{1:n}\right]\label{G_1}\\&=\min_{\delta_{n+1}}\;\mathcal{C}_{\delta_{n+1}}+\E\left[\left.\MG^N_{n+1}\lb\pi_1(n+1)\rb\right|\pi_1(n)\right],\label{G_2}
\end{align}
where \eqref{G_1} follows from the definition of ${\cal G}_{n}^N$, and \eqref{G_2} holds since ${\cal C}_{\delta_{n+1}}$ is constant given $\{X_{1:n}, \Bdelta_{1:n}\}$ and $\E\left[\left.\MG^N_{n+1}\lb\pi_1(n+1)\rb\right|X_{1:n}, \Bdelta_{1:n}\right]=\E\left[\left.\MG^N_{n+1}\lb\pi_1(n+1)\rb\right|\pi_1(n)\right]$. Substituting \eqref{G_0}-\eqref{G_2} into \eqref{Backward}, 
the backward recursion is significantly simplified to the following
 \begin{align}\label{G_BI}
{\MG}_n^N(\pi_1(n))&\!=\!
\min\!\left\{\phi\lb \pi_1(n)\rb, \min_{\delta_{n+1}}\left[\underbrace{\MC_{\delta_{n+1}}\!+\!\E\lb\left.\MG_{n+1}^N\lb\frac{\pi_1(n)\exp\lb l_{\delta_{n+1}}\rb}{\pi_0(n)+\pi_1(n)\exp\lb l_{\delta_{n+1}}\rb}\rb\right|\pi_1(n)\rb}_{\overline{\MG}_n^N(\pi_1(n),\delta_{n+1})}\right]\right\},
\end{align}
with $n=N-1, N-2, \ldots, 1, 0$. Obviously, we have 
\begin{align}\label{V0N}
\MG_0^N(\pi_1)=\MV_0^N(\pi_1) 
\end{align}
due to the definition in \eqref{def_G}. 

With the lemma below, we can further analyze the optimal stopping rule given in Lemma \ref{lemma:1}.
\begin{lemma}\label{lemma:3}
$\overline{\MG}_n^N(\pi_1(n),\delta_{n+1})$ is a concave function of $\pi_1(n)$. Moreover, the function 
\begin{align}\label{def:tildeG}
\widetilde{\cal G}_n^N(\pi_1(n))\triangleq \min_{\delta_{n+1}}\;\overline{\cal G}_n^N(\pi_1(n),\delta_{n+1})
\end{align} 
is  concave with $\widetilde{\MG}_n^N(0)>0, \widetilde{\MG}_n^N(1)>0$, for $n=0, 1, \ldots, N$.
\end{lemma}
\proof First, $\MG_N^N(\pi_1(N))=\phi(\pi_1(N))=\min\{\mu_1\pi_1(N), \mu_0(1-\pi_1(N))\}$ is concave. Second, the recursion \eqref{G_BI} suggests that, if $\MG_{n+1}^N(\pi_1(n+1))$ is concave, $\MG_{n}^N(\pi_1(n))$ is concave as well. This can be shown as follows:

Assume that $\MG_{n+1}^N(x)$ is concave, since $\frac{x\exp\lb l_{\delta_{n+1}}\rb}{1-x+x\exp\lb l_{\delta_{n+1}}\rb}$ is an increasing function of $x$ and the expectation operation preserves the concavity,  the compound function $\E\lb\left. \MG_{n+1}^N\lb\frac{x\exp\lb l_{\delta_{n+1}}\rb}{1-x+x\exp\lb l_{\delta_{n+1}}\rb}\rb\right|\pi_1(n)=x\rb$ is concave, which further leads to the concavity of $\overline{\cal G}_n^N(\pi_1(n),\delta_{n+1})$ in terms of $\pi_1(n)$; in addition, regarding $\overline{\cal G}_n^N(\pi_1(n),\delta_{n+1})$ as a series of concave functions indexed by $\delta_{n+1}$, since the point-wise minimum preserves the concavity, $\widetilde{\MG}_{n}^N(\pi_1(n))$ is a concave function; due to the same argument, the point-wise minimum of $\widetilde{\cal G}_n^N(\pi_1(n))$ and $\phi(\pi_1(n))$, i.e., $\MG_n^N(\pi_1(n))$, is concave as well. 

Therefore, by induction, we conclude that $\MG_n^N(\pi_1(n)),\; n=0, 1, \ldots, N$ are concave functions. Furthermore, from the proof above, we know that the concavity of $\MG_n^N(\pi_1(n))$ leads to the concavities of $\overline{\cal G}_n^N(\pi_1(n),\delta_{n+1})$ and $\widetilde{\MG}_{n}^N(\pi_1(n))$. Thus $\overline{\cal G}_n^N(\pi_1(n),\delta_{n+1})$ and $\widetilde{\MG}_{n}^N(\pi_1(n))$ for $n=0, 1, \ldots, N$ are concave functions. 
\endproof
Together with Lemma \ref{lemma:1}, Lemma \ref{lemma:3} reveals the following optimal stopping rule. 
\begin{lemma}
$\T^\star=\min\{n: \pi_1(n)\notin (a_n,b_n)\}$, where $a_n$ and $b_n$ are roots for 
\begin{align}\label{Lemma4_Eq}
\mu_0(1- x)= \widetilde{\MG}_{n}^N(x)\; \; \text{and}\;\; \mu_1 x= \widetilde{\MG}_{n}^N(x),
\end{align}
respectively. Moreover, $a_0<a_1<\ldots<a_N=\frac{\mu_0}{\mu_0+\mu_1}$, and $b_0>b_1>\ldots>b_N=\frac{\mu_0}{\mu_0+\mu_1}$.
\end{lemma}
\proof
See Appendix.
\endproof
Now we have obtained the optimal solution $\{\Bdelta^\star_{1:\T^\star}, D^\star_{\T^\star}, \T^\star\}$ to \eqref{P1_Bayes}, which is summarized in the theorem below. Note that we have changed the sufficient statistic $\pi_1(n)$ to its equivalent form, i.e., LLR $L_n$ to draw parallel to the well-known SPRT, and with an  abuse of notation, the selection function is also denoted as $\psi_{t+1}(L_t)$.

\begin{theorem}\label{thm1}
The optimal sequential procedure that solves \eqref{P1_Bayes} features a sequential probability ratio test with curved stopping boundary, and time-variant sensor selection strategy, i.e.,
\begin{enumerate}
\item The optimal sensor selection rule is a time-variant function of LLR: $\delta_{t+1}^\star\triangleq \psi_{t+1}(L_t)$;
\item The optimal stopping rule is in the form of a truncated SPRT, i.e.,
\begin{align}
&\T^\star=\min\{t: L_t\notin (-A_t, B_t)\},\quad \text{with}\\
&B_0>B_1>\ldots>B_N=\log\frac{\mu_0\pi_0}{\mu_1\pi_1},\; \text{and}\quad A_0>A_1>\ldots>A_N=-\log\frac{\mu_0\pi_0}{\mu_1\pi_1};
\end{align}
\item The optimal decision rule $D^\star_{\T^\star}$ decides $\mathcal{H}_0$ if $L_{\T^\star}\le -A_{\T^\star}$, and decides $\mathcal{H}_1$ if $L_{\T^\star}\ge B_{\T^\star}$.
\end{enumerate}
\end{theorem}

For the scheme given in Theorem \ref{thm1}, $\T^\star\le N$ is guaranteed by noting that $-A_N=B_N=\log \frac{\mu_0\pi_0}{\mu_1\pi_1}$, and $(-A_N, B_N)$ is an empty set. In other words, any value of $L_N$ results in stopping. In specific, $L_N\ge B_N$ gives decision $\delta_N=1$, and $L_N\le -A_N$ gives decision $\delta_N=0$. Since $L_N=-A_N=B_N=\log \frac{\mu_0\pi_0}{\mu_1\pi_1}$ holds with zero probability, the equality situation for decision can be ignored in this case. 
Theorem \ref{thm1} reveals the important structure of the optimal solution to  \eqref{P1_Bayes}, while the specific values of $A_t, B_t$ and $\psi_{t+1}(L_t)$ need to be evaluated by solving the dynamic program \eqref{G_BI}. In specific, in the posterior domain, the continuation region (i.e., the sequential test stops if the posterior goes beyond this region) and the selection region for sensor $\ell$ are given  respectively by
\begin{align}
&{\cal R}_t\triangleq \{\pi_1(t): \phi(\pi_1(t))\ge \widetilde {\cal G}_t^N(\pi_1(t))\},\label{cont_reg}\\
& {\cal D}_t^\ell\triangleq\left\{\pi_1(t): \ell=\arg\;\min_{\delta}\;{\overline{\cal G}_t^N(\pi_1(t),\delta)}\right\}, \;\ell=1, \ldots, K.\label{selection_reg}
\end{align}
Transforming ${\cal R}_t$ and ${\cal D}^\ell_t$ into the LLR domain according to \eqref{posterior}, which we denote as $\widetilde{\cal R}_t$ and $\widetilde {\cal D}^\ell_t$, then the thresholds in Theorem \ref{thm1} are evaluated as 
\begin{align}\label{thres_evaluation}
A_t=-\min \{L_t:L_t\in \widetilde{\cal R}_t\}, \quad B_t=\max \{L_t:L_t\in \widetilde{\cal R}_t\}.
\end{align}
Moreover, Lemma 3 and \eqref{selection_reg} indicate that the selection strategy boils down to finding the minimum of $K$ concave functions, i.e., $\overline{\cal G}_n^N(\pi_1(t),\delta), \; \delta=1,\ldots, K$, in the domain of posterior. Since concave functions are nicely behaved functions, the resulting selection scheme essentially partitions the domain of posterior into a finite number of intervals (assuming $K$ is finite) and assign each interval with the sensor index, whose value of $\overline{\cal G}_n^N$ is minimum within that interval. This observation suggests that, once computed offline, the sensor selection strategy can be easily stored in the fusion center. In practice, the recursion \eqref{G_BI}, the sensor selection function \eqref{selection_reg}, and  the stopping rule \eqref{cont_reg} and \eqref{thres_evaluation} are implemented by discretizing the domain of posterior $\pi_1(t)$. We summarize this procedure in Algorithm \ref{DP}, where $\bnu$ and ${\boldsymbol{L}}$ are vectors containing the discrete values of $\pi_1(t)$ and $L_t$ respectively, ${\cal G}(\bnu,t)$ and  $\psi(\bnu, t+1)$ and $\psi({\boldsymbol{L}}, t+1)$ are vectors formed by evaluating the function for each element of $\bnu$ and $\boldsymbol{L}$, representing the functions ${\cal G}_t^N(\pi_1(t))$ and $\psi_{t+1}(\pi_1(t))$, and $\psi_{t+1}(L_t)$ respectively. The expectation $\E(\cdot)=\pi_0\E(\cdot|{\cal H}_0)+\pi_1\E(\cdot|{\cal H}_1)$ therein is taken w.r.t. the distribution of random sample $X$, and is evaluated by numerical integration.  The output $\psi({\boldsymbol{L}}, t+1), t=0, 1, \ldots, N-1$ (i.e., a sequence of vectors) and $\{A(t), B(t)\}$ give the selection function and decision thresholds respectively, and ${\cal G}(\pi_1, 0)$ gives the optimal cost ${\cal G}_0^N(\pi_1)$ (or equivalently, ${\cal V}_0^N(\pi_1)$), which will be used in Section IV.
\begin{algorithm}
\caption{\bf : Procedure for computing $A_t, B_t$ and $\psi_{t+1}(L_t)$ in Theorem \ref{thm1}}
\begin{algorithmic}[1]\label{DP}
\STATE {\bf Input:} $N, \pi_1, \mu_0, \mu_1, \{\lambda_j\}_{j\in\Omega}$, the distributions of $X$ under ${\cal H}_0$ and ${\cal H}_1$\\
\STATE {\bf Initialization:} 
\\${\cal G}(\bnu,N)\leftarrow \min\lb\mu_1\bnu,\mu_0(1-\bnu)\rb$, $\psi(\bnu,N) \leftarrow 0$, $\boldsymbol{L}\leftarrow \log \frac{\pi_0\boldsymbol\nu}{\pi_1(1-\boldsymbol\nu)}$
\STATE {\bf for } $t=N-1$ to $0$ {\bf do}
\STATE Evaluate selection function at $t+1$: \\$\psi(\bnu,t+1)\leftarrow\arg \min_{\delta}\left\{\mathcal{C}_{\delta}+\E\left[ {\cal G}(\frac{\bnu e^{l_{\delta}(X)}}{1-\bnu+\bnu e^{l_{\delta}(X)}},t+1)\right]\right\}$
\STATE Update ``cost-to-go'': \\${\cal G}(\bnu,t)\leftarrow \min\left\{\min\lb\mu_1\bnu,\mu_0(1-\bnu)\rb, \mathcal{C}_{\psi(\bnu,t+1)}+\E\left[ {\cal G}(\frac{\bnu e^{l_{\psi(\pi,t+1)}(X)}}{1-\bnu+\bnu e^{l_{\psi(\pi,t+1)}(X)}},t+1)\right]\right\}$
\STATE Evaluate stopping thresholds:\\
$a(t)\leftarrow \min\left\{\nu\in \bnu: \min\lb\mu_1\nu,\mu_0(1-\nu)\rb\ge\mathcal{C}_{\psi(\nu,t+1)}+\E\left[ {\cal G}(\frac{\nu e^{l_{\psi(\nu,t+1)}(X)}}{1-\nu+\nu e^{l_{\psi(\nu,t+1)}(X)}},t+1)\right]\right\}$\\
$b(t)\leftarrow \max\left\{\nu\in\bnu: \min\lb\mu_1\nu,\mu_0(1-\nu)\rb\ge\mathcal{C}_{\psi(\nu,t+1)}+\E\left[ {\cal G}(\frac{\nu e^{l_{\psi(\nu,t+1)}(X)}}{1-\nu+\nu e^{l_{\psi(\nu,t+1)}(X)}},t+1)\right]\right\}$
\STATE Transform to the domain of LLR:\\
$A(t)\leftarrow -\log\frac{\pi_0 a(t)}{\pi_1(1-a(t))}$\\
$B(t)\leftarrow \log\frac{\pi_0 b(t)}{\pi_1(1-b(t))}$\\
$\psi({\boldsymbol{L}},t+1)\leftarrow \psi(\frac{\pi_1e^{\boldsymbol{L}}}{\pi_0+\pi_1e^{\boldsymbol{L}}},t+1)$ (which is evaluated in step 4)\\
\STATE {\bf end}\\
\STATE {\bf Output:}\\ ${\cal G}(\pi_1, 0)$, $\psi({\boldsymbol{L}}, t+1), A(t), B(t)$ for $t=0, 1, \ldots, N$
\end{algorithmic}
\end{algorithm}

\subsection{Infinite-Horizon Solution to the Bayes Problem}

Next, by building on the finite-horizon results developed in the last subsection, we consider the infinite-horizon version of the problem in \eqref{P1_Bayes}. 

The essential step of bridging the two problems is to show that the finite-horizon case approaches the infinite-horizon case as $N\to \infty$ \cite{Ferguson_book,SeqA_book,Xiaoou15}. Then the results in  the last subsection can be readily generalized to the infinite-horizon scenario. Defining the optimal cost of the infinite-horizon Bayesian problem:
\begin{align}\label{P1_Bayes_InfH}
\widetilde{\MV}(\pi_1)&\triangleq \min_{\{\T, D_\T, \Bdelta_{1:\T}\}} \;\MR\lb\Bdelta_{1:\T}, D_\T, \T\rb
\end{align}
where $\pi_1$ is the prior on $\mathcal{H}_1$. First, note that the optimal decision  function derived in \eqref{opt_decision} is independent of the horizon limit, thus $D^\star_\T$ in \eqref{DecRule} can be substituted into \eqref{P1_Bayes_InfH}, which gives the similar optimal stopping problem as that in \eqref{Opt_stopping}:
\begin{align}
\widetilde{\MV}(\pi_1)=\min_{\{\T, \Bdelta_{1:\T}\}\in {\cal A}_0^\infty} \E\lb\sum_{t=1}^\T{\cal C}_{\delta_t}+\phi(\pi_1(\T))\rb.
\end{align}
Recalling that ${\cal V}_0^N(\pi_1)= \min_{\{\Bdelta_{1:\T}, D_\T, \T \}, \T\le N}{\cal R}\lb \Bdelta_{1:\T}, D_\T, \T\rb$ according to \eqref{V_n_N}, we have the following lemma.
\begin{lemma}\label{lemma:FtoInF}
$\lim_{N\to \infty}\MV_0^N(\pi_1)=\widetilde{\MV}(\pi_1)$ for all $\pi_1\in [0,1]$.
\end{lemma}
\proof
Let $\{\Bdelta^\star_{1:{\T^\star}}, D^\star_{\T^\star}, \T^\star\}$ be the optimal solution to the infinite-horizon problem \eqref{P1_Bayes_InfH}. Define the auxiliary procedure $\{\Bdelta^\star_{1:{\widehat{\T}_N}}, D^\star_{\widehat{\T}_N}, \widehat{\T}_N\}$ where $\widehat{\T}_N=\min\{\T^\star, N\}$, then we have
\begin{align}
&\MR\lb \Bdelta^\star_{1:{\widehat{\T}_N}}, D^\star_{\widehat{\T}_N}, \widehat{\T}_N\rb-\MR\lb \Bdelta^\star_{1:{{\T}^\star}}, D^\star_{{\T}^\star}, {\T}^\star\rb\nonumber\\=&\;\E\lb\mathbbm{1}_{\{\T^\star\ge N\}}\lb\phi\lb \pi_1(\widehat{\T}_N)\rb-\phi\lb \pi_1(\T^\star)\rb-\sum_{t=N+1}^\infty\MC_{\delta_t}\rb\rb\nonumber\\\le &\; \;\E\lb\mathbbm{1}_{\{\T^\star\ge N\}}\lb\phi\lb \pi_1(\widehat{\T}_N)\rb\rb\rb
\label{fin-to-inf_1}\\=&\;\E\lb\phi\lb\pi_1(N)\rb \mathbbm{1}_{\{\widehat{\T}_N=N\}}\rb,\label{fin-to-inf_2}
\end{align}
where \eqref{fin-to-inf_1} follows from the fact that $\phi(\pi_1(\T^\star))$ and ${\cal C}_{\delta_t}$ are positive, and \eqref{fin-to-inf_2} is true because $\widehat{\T}_N=N$ holds with probability one given that $\T^\star\ge N$ due to the definition of $\widehat{\T}_N$. 
Using \eqref{fin-to-inf_2} and the fact that ${\cal V}_0^N(\pi_1)$ is the optimal cost for all $\T\le N$ whereas $\{\Bdelta^\star_{1:{\widehat{\T}_N}}, D^\star_{\widehat{\T}_N}, \widehat{\T}_N\}$ is a constructed scheme for $\T\le N$,  we arrive at the following inequalities
\begin{align}
\MV_0^N(\pi_1)\le \MR\lb \Bdelta^\star_{1:{\widehat{\T}_N}}, D^\star_{\widehat{\T}_N}, \widehat{\T}_N\rb\le \MR\lb \Bdelta^\star_{1:{{\T}^\star}}, D^\star_{{\T}^\star}, {\T}^\star\rb+ \E\lb\phi\lb\pi_1(N)\rb \mathbbm{1}_{\{\widehat{\T}_N=N\}}\rb.\label{limit}
\end{align}
By the strong law of large number, we know that $L_N\to \infty, \;\text{a.s.}$ as $N\to \infty$, thus $\phi\lb\pi_1(N)\rb=\min\{\mu_0\pi_0(N),\mu_1\pi_1(N)\}\to 0\; \text{a.s.}$ as $N\to\infty$ \cite{Ferguson_book}. Taking $N\to \infty$ on both sides of \eqref{limit}, we have
\begin{align}\label{lemma5_1}
\lim_{N\to \infty}\MV_0^N(\pi_1)\le \MR\lb \Bdelta^\star_{1:{{\T}^\star}}, D^\star_{{\T}^\star}, {\T}^\star\rb=\widetilde{\cal V}(\pi_1).
\end{align}
On the other hand, ${\cal V}_0^N(\pi_1)\ge \MR\lb \Bdelta^\star_{1:{{\T}^\star}}, D^\star_{{\T}^\star}, {\T}^\star\rb$, since ${\cal V}_0^N(\pi_1)$ is the minimal  cost for the finite-horizon problem, i.e., $\T\le N$, whereas $\MR\lb \Bdelta^\star_{1:{{\T}^\star}}, D^\star_{{\T}^\star}, {\T}^\star\rb$ is the minimal cost for the  infinite-horizon problem, where no bound on $\T$ is imposed. Thus, we have $\lim_{N\to \infty} {\cal V}_0^N(\pi_1)\ge \MR\lb \Bdelta^\star_{1:{{\T}^\star}}, D^\star_{{\T}^\star}, {\T}^\star\rb=\widetilde{\cal V}(\pi_1)$ that, together with \eqref{lemma5_1}, completes the proof.
\endproof

Meanwhile, in the finite-horizon solution \eqref{G_BI}, since $\MG_n^N(\pi_1(n))$ is a function of the homogenous Markov chain $\pi_1(n)$, we have $\MG_n^N(x)=\MG_0^{N-n}(x)=\MV_0^{N-n}(x)$.  The first equality follows from the homogeneity property, and second equality follows from definitions. Therefore, the backward induction \eqref{G_BI} can be equivalently expressed as the recursion
\begin{align}
&{\MV}_0^{N-n}(x)=
\min\left\{\phi\lb x\rb, \min_{\delta_{n+1}}\left[\MC_{\delta_{n+1}}+\E\lb\MV_{0}^{N-n-1}\lb\frac{x\exp\lb l_{\delta_{n+1}}\rb}{1-x+x\exp\lb l_{\delta_{n+1}}\rb}\rb\rb\right]\right\},
\end{align}
\ignore{
 \begin{align}
{\MV}_0^{N}(x)&=
\min\left\{\phi\lb x\rb, \min_{\delta}\left[\MC_{\delta}+\E\lb\left.\MV_{0}^{N-1}\lb\frac{x\exp\lb l_{\delta}\rb}{1-x+x\exp\lb l_{\delta}\rb}\rb\right|x\rb\right]\right\},
\end{align}}
with $\MV_0^0(x)=\phi(x)$.
By letting $N\to \infty$, and invoking Lemma \ref{lemma:FtoInF}, we arrive at
\begin{align}\label{Bellman}
&\widetilde{\MV}(x)=
\min\left\{\phi\lb x\rb, \min_{\delta}\left[\MC_{\delta}+\E\lb\widetilde{\MV}\lb\frac{x\exp\lb l_{\delta}\rb}{1-x+x\exp\lb l_{\delta}\rb}\rb\rb\right]\right\}.
\end{align}
This is the Bellman equation for the infinite-horizon Bayesian problem \eqref{P1_Bayes_InfH}. Note that, thanks to Lemma \ref{lemma:FtoInF}, $\widetilde{\cal V}\lb x\rb$ preserves the concavity of $\MV_0^N$. Therefore, \eqref{Bellman} reveals that the stopping boundaries under infinite-horizon are constants. Moreover, the sensor selection function $\delta_{t+1}$ depends only on the  posterior/LLR, and is independent of time. We summarize the optimal solution to the infinite-horizon problem in the theorem below. 
\begin{theorem}\label{thm2}
The optimal procedure that solves \eqref{P1_Bayes} features an SPRT with stationary sensor selection strategy, i.e.,
\begin{enumerate}
\item The optimal sensor selection rule is a time-invariant function of the likelihood raito, i.e., $\delta^\star_{t+1}=\psi(L_t)$. 
\item The stopping rule is in the form of the  SPRT $\T^\star=\min\{t: L_t\notin (-A, B)\}$.
\item The optimal decision rule $D^\star_{\T^\star}$ decides $\mathcal{H}_0$ if $L_{\T^\star}\le -A$, and decides $\mathcal{H}_1$ if $L_{\T^\star}\ge B$.
\end{enumerate}
The function $\psi(L_t)$ and the thresholds $A, B$ can be evaluated numerically by solving the Bellman equation \eqref{Bellman}.
\end{theorem}
The proof for Theorem \ref{thm2} follows similarly to that of Theorem \ref{thm1} by using the Bellman equation \eqref{Bellman}. In brief, $\widetilde {\cal V}(x)$ and ${\cal E}(x) \triangleq \min_\delta \left[{\cal C}_\delta+\E\lb \widetilde{\cal V}\lb \frac{x \exp(l_\delta)}{\lb 1-x+x\exp(l_\delta)\rb}\rb\rb\right]$ can be proved to be  concave functions with ${\cal E}(0)>0$ and ${\cal E}(1)>0$ by letting $N\to \infty$ in Lemma 3; then the operation $\min_\delta$ in ${\cal E}(x)$ indicates that the selection rule is a time-invariant function of the posterior, leading to Theorem 2-(1); moreover, analogous to \eqref{Lemma4_Eq} in Lemma 3, the stopping thresholds are given by the roots for $\mu_0(1-x)={\cal E}(x)$ and $\mu_1x={\cal E}(x)$ which are constants, leading to Theorem 2-(2). The key difference here is that ${\cal E}(x)$ is independent of $n$ in contrast with $\widetilde {\cal G}_n^N(x)$ in the proof of Theorem 1. Interestingly, Theorem \ref{thm2} implies that the stopping thresholds and selection strategy of the infinite-horizon Bayesian problem converge to a sequential procedure that, in essence, is a combination of the SPRT and stationary sensor selection function $\psi(L_t)$. Several approaches are available to solve the Bellman equation for $\psi(L_t)$ and $A, B$. In this work, by  virtue of Lemma \ref{lemma:FtoInF}, we solve a finite-horizon problem with sufficiently  large $N$ to approximately obtain them, which will be explained in Section IV.
\subsection{Optimal Solution to the Usage-Constrained Problem}

Now that the Bayesian optimal stopping problem is solved in the previous subsections, we are ready to establish the optimal sequential procedure for \eqref{P1} as follows. 
\begin{theorem}\label{thm3}
Let $\bmu\triangleq [\mu_0, \mu_1]$ be chosen such that the reliability constraints are satisfied with equalities; let $\bla\triangleq \{\lambda_j\}_{j\in \Omega}$ be chosen such that all usage constraints are satisfied, and moreover, the usage constraints for the sensors in $\Omega_c\triangleq\{\ell:\lambda_\ell>0\}$ are satisfied with equalities. Then the optimal sequential procedure given by Theorems \ref{thm1} and \ref{thm2} give the optimal triplets $\{\T^\star, D^\star_{\T^\star}, \Bdelta^\star_{1:\T^\star}\}$ that solve the constrained problem \eqref{P1} in finite-horizon and infinite-horizon scenarios, respectively.
\end{theorem}
\proof
The proofs are the same for finite-horizon and inifite-horizon problems, thus we only show the latter for conciseness. 

Considering the results in Section III-A\&B, we have $\MR\lb \Bdelta_{1:\T}, D_\T, \T\rb\ge \MR\lb {\Bdelta^\star}_{1:{\T^\star}}, D^\star_{\T^\star}, \T^\star\rb$ for any procedure $\{\Bdelta_{1:\T}, D_\T, \T\}$. That is
\begin{align}
\E\T+&\mu_0\pi_0\Prob_0\lb D_\T=1\rb+\mu_1\pi_1\Prob_1\lb D_\T=0\rb +\sum_{\ell\in\Omega_c}\lambda_\ell \E\lb\sum_{t=1}^\T\mathbbm{1}_{\{\delta_t=\ell\}}\rb\nonumber\\&\ge\E\T^\star+\mu_0\pi_0\Prob_0\lb D^\star_{\T^\star}=1\rb+\mu_1\pi_1\Prob_1\lb D^\star_{\T^\star}=0\rb+\sum_{\ell\in\Omega_c}\lambda_\ell\E\lb\sum_{t=1}^{\T^\star}\mathbbm{1}_{\{\delta^\star_t=\ell\}}\rb\nonumber\\&=\E\T^\star+\mu_0\pi_0\alpha+\mu_1\pi_1\beta+\sum_{\ell\in\Omega_c} \lambda_\ell T^\ell.
\end{align}
Note that $\mu_0\ge 0$, $\mu_1\ge 0$ and $\lambda_\ell> 0$ for $\ell\in\Omega_c$, thus $\E\T\ge \E\T^\star$ must hold true for any procedure $\{\Bdelta_{1:\T}, D_\T, \T\}\in {\bf C}\lb \alpha, \beta, \{T^\ell\}_{\ell\in \Omega}\rb$.
\endproof
The insight for Theorem \ref{thm3} is intuitive. The sensors in $\Omega_c$ (referred to as the effective set henceforth) will be overused without imposing the constraint, thus additional sampling cost $\lambda_\ell>0$ is assigned to penalize  their usages (recall the definition of $\mathcal{C}_{\delta_t}$ in \eqref{Bayes_obj}). Nevertheless, in order to optimize the test performance, they should be used at full capacity, i.e., usage constraints are satisfied with equalities. Section IV will address how we obtain $\Omega_c$ from a general set $\Omega$ that are under usage constraints in the formulation \eqref{P1}.


Next, we investigate the performance of the optimal sequential procedure under infinite-horizon. The challenge stems from the fact that random samples are no longer i.i.d., and the typical method based on Wald's identity fails to given valid performance analysis. However, by capitalizing on the optimal structures revealed in Theorems 2 and 3, we can derive an insightful bound to approximately characterize the performance. Define the Kullback-Leibler divergence (KLD):
\begin{align}
\MD^\ell_i\lb f^{\ell}_i||f^{\ell}_j\rb\triangleq \E_i\lb\log\frac{f_i^\ell(X)}{f_j^\ell(X)}\rb.
\end{align}
\begin{proposition}\label{cor1}
Based on the Wald's approximation \cite{SeqA_book} (i.e., $L_{\T^\star}\approx -A$  given $D^\star_{\T^\star}=0$ or $L_{\T^\star}\approx B$ given $D^\star_{\T^\star}=1$), the expected sample size for the optimal procedure for the infinite-horizon problem of \eqref{P1} is lower bounded by
\begin{align}\label{Bound}
&\E\T^\star\ge\nonumber\\
&\pi_0\frac{\MD\lb \alpha||1-\beta\rb}{\max_{\ell\in\overline{\Omega}_c} \MD_0^\ell}+\pi_1\frac{\MD\lb1-\beta||\alpha\rb}{\max_{\ell\in\overline{\Omega}_c} \MD_1^\ell}-\sum_{\ell\in\Omega_c}\lb\max\left\{\frac{\MD^\ell_1}{\max_{\ell\in\overline{\Omega}_c} \MD_1^\ell},\frac{\MD^\ell_0}{\max_{\ell\in\overline{\Omega}_c} \MD_0^\ell}\right\}-1\rb T^\ell,
\end{align}
where $\MD\lb p||q\rb \triangleq p\log \frac{p}{q}+(1-p)\log\frac{1-p}{1-q}$ is the KLD of binary distribuitons, and $\overline{\Omega}_c\triangleq \Pi\backslash_ {\Omega_c}$ contains all sensors except those in $\Omega_c$.
\end{proposition}
\proof
See Appendix.
\endproof
The performance characterization agrees with intuition. The first two terms on right-hand side of \eqref{Bound} characterize the asymptotic performance of the optimal sequential procedure as $\alpha$ and $\beta$ go to zero, or $\MD\lb \alpha||1-\beta\rb$ and $\MD\lb1-\beta||\alpha\rb$ go to infinity. It is seen that the asymptotic expected sample size is determined by the KLDs of the sensors in $\overline{\Omega}_c$, i.e., the free sensors that do not reach their full usage. This result is consistent with that in \cite{Javidi10}, where all sensors are constraint-free.  Meanwhile, the third term on the right-hand side of \eqref{Bound} accounts for the effect of the fully used sensors, which depends on their KLDs compared to that of the free  sensors. If $\max\left\{\frac{\MD^\ell_1}{\max_{\ell\in\overline{\Omega}_c} \MD_1^\ell},\frac{\MD^\ell_0}{\max_{\ell\in\overline{\Omega}_c} \MD_0^\ell}\right\}> 1$, then sensor $\ell$ decreases the expected sample size due to its larger KLDs; otherwise, sensor $\ell$ increases the expected sample size.
\section{Parameters Design for the Optimal Sequential Test}\label{sec:algorithm}

In previous sections, we derived the optimal solutions to \eqref{P1} under both finite-horizon and infinite-horizon setups, given that $\bmu$ and $\bla$ are set to satisfy certain conditions as given in Theorem \ref{thm3}. These multipliers determine the parameters in the optimal sequential test and selection function, i.e., $A_t, B_t, \psi_{t+1}(L_t)$ for finite-horizon, and $A, B, \psi(L_t)$ for infinite-horizon. In practice, one can choose the multipliers by manually refining their values according to the simulation results; however, it is not an efficient approach, especially when the number of constraints is large.  In this section, we propose a systematic approach to approximately evaluate the multipliers, which involves minimizing a concave function. 

By drawing on the idea of the recent work \cite{Fauss15}, we evaluate the multipliers by introducing the dual problem of \eqref{P1}:
\begin{align}\label{DualProb}
\max_{\{\bla, \bmu\}\in\mathbb{R}^+} \min_{\{\Bdelta_{1:\T}, D_\T, \T\}}\ML(\{\Bdelta_{1:\T}, D_\T, \T\}, \bla, \bmu),
\end{align}
where the Lagrangian admits
\begin{align}
\ML(\{\Bdelta_1^\T, D_\T, \T\}, \bla, \bmu)\triangleq \E\T&+\mu_0\pi_0\lb\Prob_0\lb{D_\T=1}\rb-\alpha\rb\nonumber\\&+\mu_1\pi_1\lb\Prob_1\lb{D_\T=0}\rb-\beta\rb+\sum_{\ell\in\Omega}\lambda_\ell \lb \sum_{t=1}^\T\mathbbm{1}_{\{\delta_t=\ell\}}-T^\ell\rb\nonumber\\
=\MR&(\Bdelta_{1:\T}, D_\T, \T)-\mu_0\pi_0\alpha-\mu_1\pi_1\beta-\sum_{\ell\in\Omega}\lambda_\ell T^\ell.
\end{align}
The reason is that if there exist multipliers such that the constraints hold as equalities, they must reside in the saddle point as expressed in \eqref{DualProb}.  

We first begin with the $N$-horizon problem. Since the Bayesian problem is solved in Section III, \eqref{DualProb} becomes 
\begin{align}\label{Dual2}
\max_{\{\bla, \bmu\}\in \mathbb{R}^+}\widetilde{\ML}_N(\bla, \bmu)&\triangleq \underbrace{\min_{\{D, \T, \Bdelta_{1:\T}\}}\; \E\lb \sum_{t=1}^\T \MC_{\delta_t}+\mu\lb D_\T, \MH\rb\rb}_{{\MV}_0^N(\pi_1, \bla, \bmu)}-\sum_{\ell\in\Omega}\lambda_\ell T^\ell-\mu_0\pi_0\alpha-\mu_1\pi_1\beta,
\end{align}
where $\widetilde{\ML}_N(\bla, \bmu)$ is a concave function of $\bla$ and $\bmu$. Note that ${\cal V}_0^N(\pi_1, \bla, \bmu)$ is the same function as defined in \eqref{V0N} while we explicitly  show the variables $\bla$ and $\bmu$ here for clarity. 

Note that \eqref{Dual2} is a constrained concave problem that still requires complex solving process, for example, the interior-point method \cite{Boyd_book}. In this work, we propose a simple procedure based on gradient ascent. In brief, we first assume that the effective set of constraints $\Omega_c$ is known,  based on which, \eqref{Dual2} can be recast into an  unconstrained optimization problem; we then give the scheme for evaluating $\Omega_c$. The detailed procedure includes the following steps:
\begin{itemize}
\item Given any $\Omega_c$, it is known that the optimal multipliers $\mu_0>0$, $\mu_1>0$, $\lambda_j> 0$ for $j\in \Omega_c$ and $\lambda_j=0$ for $j\in \overline{\Omega}_c$ (cf. Theorem \ref{thm3}). Consequently, the original problem \eqref{Dual2} can be reduced to  an unconstrained problem by removing $\lambda_j, \;j\in\overline{\Omega}_c$:
\begin{align}\label{Dual2_unconstrained}
\max_{\bla_{\Omega_c}, \bmu}\widetilde{\ML}_N(\bla_{\Omega_c}, \bmu)&\triangleq {{\MV}_0^N(\pi_1, \bla_{\Omega_c}, \bmu)}-\sum_{\ell\in\Omega_c}\lambda_\ell T^\ell-\mu_0\pi_0\alpha-\mu_1\pi_1\beta,
\end{align}
with $\bla_{\Omega_c}\triangleq \{\lambda_j\}_{j\in \Omega_c}$,  since the optimal values of  $\lambda_j, \; j\in\Omega_c$ and $\bmu$ reside in the interior of the positiveness constraint. Now \eqref{Dual2_unconstrained} can be solved with the gradient ascent algorithm. To this end, note that $\MV_0^N(\pi_1, \bla_{\Omega_c}, \bmu)$ can be obtained efficiently given any value of the variables $\bmu, \bla_{\Omega_c}$ through the dynamic programming \eqref{G_BI}, i.e., Algorithm 1. This allows us to approximate the gradients at the $t$th iteration by using small shifts $\Delta_\bla$ and $\Delta_\bmu$ for $\bla_{\Omega_c}$ and $\bmu$ respectively.
\ignore{\begin{align*}
&\nabla_\bla\widetilde{\ML}_N(\bla^{(t)},\bmu^{(t)})=\nabla_\bla\MV^N(\bla^{(t)},\bmu^{(t)})-[N_1;\ldots; N_{|\Omega_c|}],\\
&\nabla_\bmu\widetilde{\ML}_N(\bla^{(t)},\bmu^{(t)})=\nabla_\bla\MV^N(\bla^{(t)},\bmu^{(t)})-[\alpha; \beta],
\end{align*}
with
\begin{align*}
\nabla_\bla\MV_0^N(\bla^{(t)},\bmu^{(t)})\triangleq&\left.\frac{\partial \MV^N(\pi, \bla, \bmu^{(t)})}{\partial \bla}\right|_{\bla=\bla^{(t)}}\\\approx \Big[&\frac{\MV^N(\pi, [\la^{(t)}_1+\Delta_\la, \la^{(t)}_2, \ldots, \la^{(t)}_L], \bmu^{(t)})-\MV^N(\pi, \bla^{(t)}, \bmu^{(t)})}{\Delta_\la};\\&\frac{\MV^N(\pi, [\la^{(t)}_1, \la^{(t)}_2+\Delta_\la, \ldots, \la^{(t)}_L], \bmu^{(t)})-\MV^N(\pi, \bla^{(t)}, \bmu^{(t)})}{\Delta_\la};\nonumber\\&\quad\vdots\nonumber\\& \frac{\MV^N(\pi, [\la^{(t)}_1, \ldots, \la^{(t)}_L+\Delta_\la], \bmu^{(t)})-\MV^N(\pi, \bla^{(t)}, \bmu^{(t)})}{\Delta_\la}\Big],\\
\nabla_\bmu\MV_0^N(\bla^{(t)}, \bmu^{(t)})\triangleq &\left.\frac{\partial \MV^N(\pi, \bla^{(t)}, \bmu)}{\partial \bmu}\right|_{\bmu=\bmu^{(t)}}\\\approx \Big[&\frac{\MV^N(\pi, \bla^{(t)}, [\mu_0^{(t)}+\Delta_\mu; \mu^{(t)}_1])-\MV^N(\pi, \bla^{(t)}, \bmu^{(t)})}{\Delta_\mu};\nonumber\\& \frac{\MV^N(\pi, \bla^{(t)}, [\mu_0^{(t)}; \mu_1^{(t)}+\Delta_\mu])-\MV^N(\pi, \bla^{(t)}, \bmu^{(t)})}{\Delta_\mu}\Big].
\end{align*}}
Moreover, since $\bmu$ and $\bla_{\Omega_c}$ are typically at different scales, for example, $\bmu$ are usually in the order of hundreds, while $\bla_{\Omega_c}$ are  fractional numbers, we apply the alternating minimization to speed up the convergence. Algorithm \ref{Alg_la1} summarizes the procedure for evaluating the multipliers and the resulting parameters (i.e., $A_t, B_t, \psi_{t+1}(L_t)$) for the finite-$N$ optimal sequential test, where $\text{Alg}_1(\cdot)$ invokes Algorithm 1. In addition, $p_t$ and $q_t$ are step-sizes obtained by backtracking line search \cite{Boyd_book}, $\bmu_\text{int}, \bla_\text{int}$ are initial values to begin the iterations.

\item To obtain the effective set $\Omega_c$, we add an outer iteration to Algorithm \ref{Alg_la1}. In particular,
\begin{enumerate}
\item Begin with an empty set of effective usage constraints (i.e., $\Omega_c=\emptyset$).
\item Solve the problem 
\begin{align}\label{temp_problem}
\min_{\{\Bdelta_{1:\T}, D_\T, \T\}\in{\bf C}_N\lb \alpha, \beta, \{T^\ell\}_{\ell\in \Omega_c}\rb}\;\;  \E \T.
\end{align}
\item Evaluate the sensor usages based on the solution to \eqref{temp_problem}, and find the set of sensors in $\Omega$ whose constraints are violated (denoted as $\Lambda$). Update the effective set $\Omega_c\leftarrow \Omega_c\cup\Lambda$.
\item Go to step 2) and solve \eqref{temp_problem} for the updated $\Omega_c$.
\end{enumerate}
This loop of 2)-4) continues until no inequality constraints are violated. Upon termination, $\Omega_c$ is effective set of constraints, whose associated multipliers are positive, whereas the rest of constraints are naturally satisfied with zero multipliers. 
\end{itemize}
\begin{algorithm}
\caption{\bf : Procedure for solving \eqref{Dual2_unconstrained}}
\begin{algorithmic}[1]\label{Alg_la1}
\STATE Initialization: $t\leftarrow 0, \bmu^{(0)}\leftarrow{\bmu_\text{int}}, \bla_{\Omega_c}^{(0)}\leftarrow\bla_\text{int}$
\STATE {\bf while} $\lVert\nabla_\bla\widetilde{\ML}^N(\bla_{\Omega_c}^{(t)},\bmu^{(t)})\rVert_2>\epsilon_0\; \text{or}\;\lVert\nabla_\bla\widetilde{\ML}^N(\bla_{\Omega_c}^{(t)},\bmu^{(t)})\rVert_2>\epsilon_1$ {\bf do}\\
\;\;{\bf update $\bmu$:}\\
\STATE \quad ${\cal V}_0^N(\pi_1, \bla_{\Omega_c}^{(t)}, \bmu^{(t)})\leftarrow{\cal G}(\pi_1,0)\leftarrow\text{Alg}_1(\pi_1, \bla_{\Omega_c}^{(t)}, \bmu^{(t)})$
\STATE \quad ${\cal V}_0^N(\pi_1, \bla_{\Omega_c}^{(t)}, \bmu^{(t)}+\Delta_\bmu)\leftarrow{\cal G}(\pi_1,0)\leftarrow\text{Alg}_1(\pi_1, \bla_{\Omega_c}^{(t)}, \bmu^{(t)}+\Delta_\bmu)$ 
\STATE \quad Evaluate $\widetilde {\cal L}_N(\bla_{\Omega_c}^{(t)}, \bmu^{(t)})$ and $\widetilde {\cal L}_N(\bla_{\Omega_c}^{(t)}, \bmu^{(t)}+\Delta_\bmu)$ by its definition in \eqref{Dual2_unconstrained}
\STATE \quad Approximate the gradient $\nabla_{\bmu}\widetilde{\ML}_N(\pi, \bla_{\Omega_c}^{(t)}, \bmu^{(t)})$ 
\STATE \quad Update $\bmu^{(t+1)}=\bmu^{(t)}+p_t\nabla_{\bmu}\widetilde{\ML}_N(\pi, \bla_{\Omega_c}^{(t)}, \bmu^{(t)})$, where $p_t$ is the step-size computed by \\ \quad backtracking line search\\
\;\;{\bf update $\bla$:}\\
\STATE \quad ${\cal V}_0^N(\pi_1, \bla_{\Omega_c}^{(t)}, \bmu^{(t+1)})\leftarrow{\cal G}(\pi_1,0)\leftarrow\text{Alg}_1(\pi_1, \bla_{\Omega_c}^{(t)}, \bmu^{(t+1)})$
\STATE \quad ${\cal V}_0^N(\pi_1, \bla_{\Omega_c}^{(t)}+\Delta_\bla, \bmu^{(t+1)})\leftarrow{\cal G}(\pi_1,0)\leftarrow\text{Alg}_1(\pi_1, \bla_{\Omega_c}^{(t)}+\Delta_\bla, \bmu^{(t+1)})$ 
\STATE \quad Evaluate $\widetilde {\cal L}_N(\pi_1, \bla_{\Omega_c}^{(t)}, \bmu^{(t+1)})$ and $\widetilde {\cal L}_N(\pi_1, \bla_{\Omega_c}^{(t)}+\Delta_\bla, \bmu^{(t+1)})$ by its definition in \eqref{Dual2_unconstrained}
\STATE \quad Approximate the gradient $\nabla_{\bla}\widetilde{\ML}_N(\pi, \bla_{\Omega_c}^{(t)}, \bmu^{(t+1)})$
\STATE \quad Update $\bla_{\Omega_c}^{(t+1)}=\bla_{\Omega_c}^{(t)}+q_t\nabla_{\bla}\widetilde{\ML}_N(\pi, \bla_{\Omega_c}^{(t)}, \bmu^{(t+1)})$ where $q_t$ is the step-size computed by \\ \quad backtracking line search
\STATE \quad $t \leftarrow t+1$
\STATE {\bf end while}
\STATE Output:\\ $\bla_{\Omega_c}^\star\leftarrow \bla_{\Omega_c}^{(t)}$, $\bmu^\star\leftarrow \bmu^{(t)}$, 
$\left\{\psi(\boldsymbol{L}, t), A(t), B(t)\right\}_{t=0}^N\leftarrow \text{Alg}_1(\pi_1,\bla_{\Omega_c}^{\star},\bmu^{\star})$
\end{algorithmic}
\end{algorithm}
\ignore{\begin{algorithm}
\caption{\bf : Evaluating the Effective Set of Usage Constraints}
\begin{algorithmic}[1]\label{Alg_la2}
\STATE Initialization: $\Omega_c\leftarrow \emptyset,  \Lambda\leftarrow \Omega, \bla_{\Omega_c}\leftarrow {\bf 0}, \bmu\leftarrow \bmu_\text{int}$
\STATE {\bf while} $\Lambda\neq \emptyset$ {\bf do}
\STATE\quad Obtain the sequential test using Algorithm 1 with the current $\bmu$ and $\bla_{\Omega_c}$
\STATE \quad Evaluate the usages for all sensors and find $\Lambda$ to be the set of sensors that do not meet \\\quad usage constraints
\STATE \quad Update the effective set: $\Omega_c\leftarrow \Omega_c\cup \Lambda$
\STATE \quad Solve \eqref{Dual2_unconstrained} using Algorithm \ref{Alg_la1} to update $\bmu$ and $\bla_{\Omega_c}$
\STATE {\bf end while}
\end{algorithmic}
\end{algorithm}}

Next we consider the infinite-horizon scenario, whose evaluation of multipliers boils down to the following optimization problem:
\begin{align*}
&\max_{\{\bla, \bmu\}\in \mathbb{R}^+} \; \widetilde{\MV}(\pi_1, \bla, \bmu)-\mu_0\pi_0\alpha-\mu_1\pi_1\beta-\sum_{\ell\in\Omega}\lambda_\ell T^\ell\\
&\text{s.t.}\; \; \widetilde{\MV}(x, \bla, \bmu)\!=\!\min\!\left\{\!\mu_0(1-x),\mu_1x, \min_{\delta}\lb \!1\!+\!\la_{\delta}\!+\!\E\lb\widetilde{\MV}(\frac{xe^{l_\delta}}{1-x+xe^{l_\delta}}, \bla, \bmu)\rb\rb\right\}, \; x\in [0, 1].
\end{align*}
One option is to adopt the method in \cite{Fauss15} (only SPRT and $\bmu$ were of interest there), which discretizes $x, \bla, \bmu$, and recasts the above problem into a linear program. However,  this approach becomes computationally infeasible due to the high-dimensional variables in our problem. To that end, by the virtue of Lemma 5, we propose to approximate the infinite-horizon problem through finite-horizon approach \eqref{Dual2}, i.e., $\widetilde{\MV}\approx\MV_0^N$ with sufficiently large $N$.  
Moreover, we obtain the multipliers and the resulting test parameters (i.e., $A, B, \psi(L_t)$) for the optimal infinite-horizon sequential test by setting $A\leftarrow A(0)$, $B\leftarrow B(0)$, $\psi(\boldsymbol{L})\leftarrow \psi(\boldsymbol{L},1)$, where $A(0), B(0)$ and $\psi(\boldsymbol{L},1)$ are the thresholds and selection function respectively evaluated for the finite-horizon problem with large $N$.


\ignore{\begin{algorithm}
\caption{\bf : Solving for Multipliers for Infinite-Horizon Problem}
\begin{algorithmic}[1]\label{Alg_la_gen}
\STATE Initialization: $\Omega_c\leftarrow \emptyset,  \Lambda\leftarrow \Omega, \bla\leftarrow {\bf 0}, \bmu\leftarrow \bmu_\text{int}$
\STATE Choose a sufficient large $N$ (the performance bound is helpful)
\STATE Run Algorithm \ref{Alg_la1}-\ref{Alg_la2}
\STATE Output: 
\end{algorithmic}
\end{algorithm}}
\section{Numerical Results}
In this section, we provide numerical results to illustrate the theoretical findings in previous sections, and also to compare with the existing methods. Our experiments focus on the following hypotheses 
\begin{align*}
&\MH_0: X_t\sim \text{exp}\lb\eta_0^\ell\rb, \quad t=1, 2, \ldots, \quad \ell\in\{1, 2, \ldots, 4\},\\
&\MH_1: X_t\sim \text{exp}\lb\eta_1^\ell\rb, \quad t=1, 2, \ldots, \quad \ell\in\{1, 2, \ldots, 4\}. 
\end{align*}
\begin{table*}
\centering
\caption{}
  \begin{tabular}{ | c | c | c | c | c |}
    \hline
    \phantom{1} & $\eta^\ell_0$  & $\eta^\ell_1$ & $D^\ell_0$ &$D^\ell_1$ \\ \hline
    Sensor 1 & $0.5$  & $1$& $0.2692$ & $0.1739$ \\ \hline
    Sensor 2  & $1$ & $0.5$  & $0.1739$ & $0.2692$   \\ \hline    
    Sensor 3 & $0.52$  & $1$ & $0.3069$ & $0.1931$\\ \hline
    Sensor 4 & $1$  & $0.52$ & $0.1931$ & $0.3069$ \\ \hline
  \end{tabular}\label{Table:models}
\end{table*} 
In particular, the LLR at sensor $\ell$ is 
\begin{align}
l^\ell(X_t)=X_t\lb\eta^\ell_0-\eta^\ell_1\rb+\log\lb \frac{\eta^\ell_1}{\eta^\ell_0}\rb
\end{align}
and the KLDs are expressed respectively as
\begin{align}
&\MD_1^\ell=\E_0\lb l^\ell\rb= \frac{\eta^\ell_0}{\eta^\ell_1}-1-\log\lb\frac{\eta_0^\ell}{\eta^\ell_1}\rb,\\
&\MD_0^\ell=\E_0\lb -l^\ell\rb= \frac{\eta^\ell_1}{\eta^\ell_0}-1-\log\lb\frac{\eta^\ell_1}{\eta_0^\ell}\rb.
\end{align}
Table \ref{Table:models} lists the distribution parameters and KLD for each sensor. Throughout the experiment, the domain of posterior $[0, 1]$ is discretized into $8000$ points to implement Algorithm 1.  
\subsection{Finite-Horizon Scenario}

We  first consider a finite-horizon problem with sample size limit $N=100$. 

\begin{figure}
\centering
\subfigure[Unconstrained]{\includegraphics[width=0.49\textwidth]{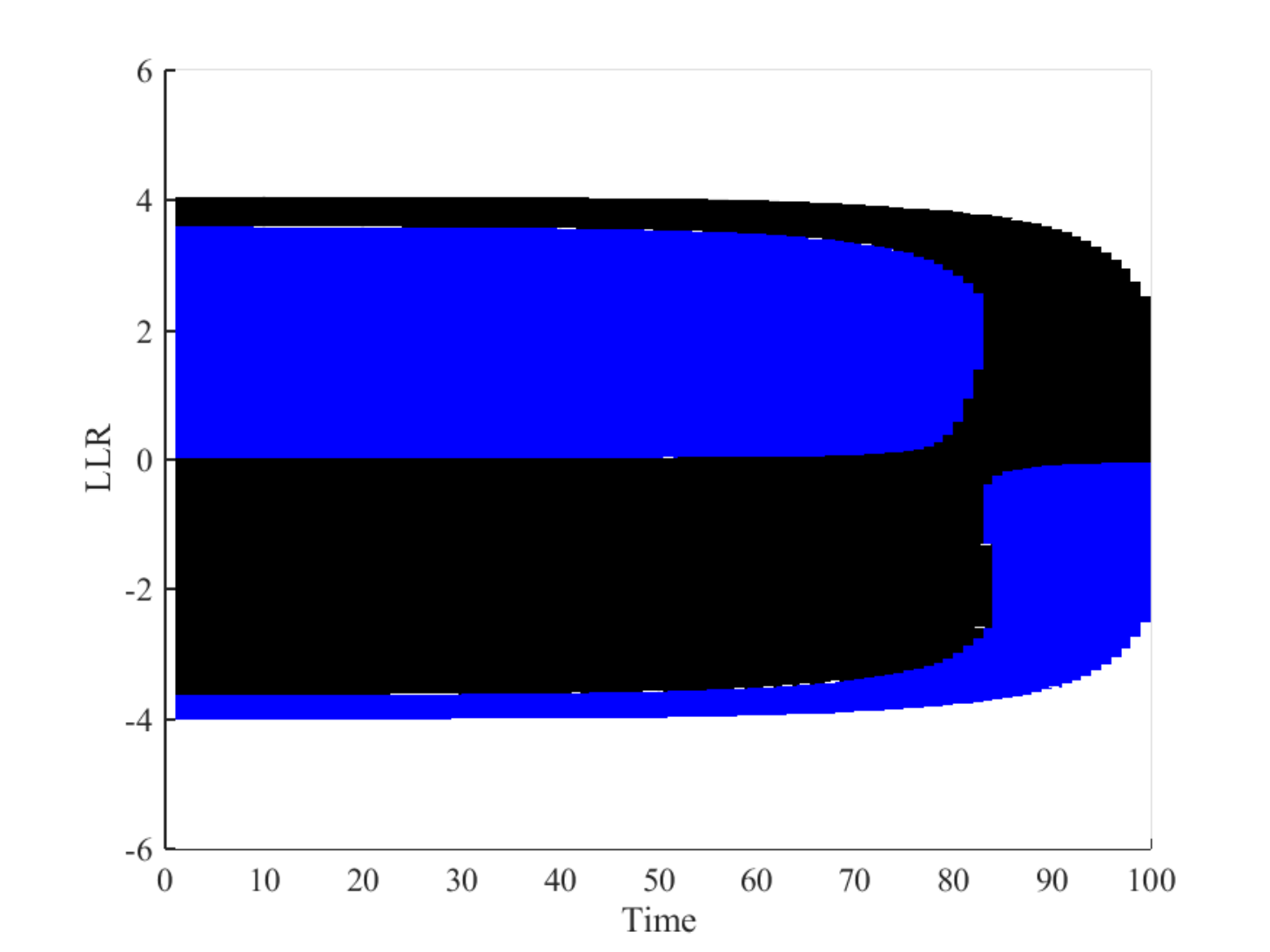}}
\subfigure[$T^1=7, T^2=7$]{\includegraphics[width=0.49\textwidth]{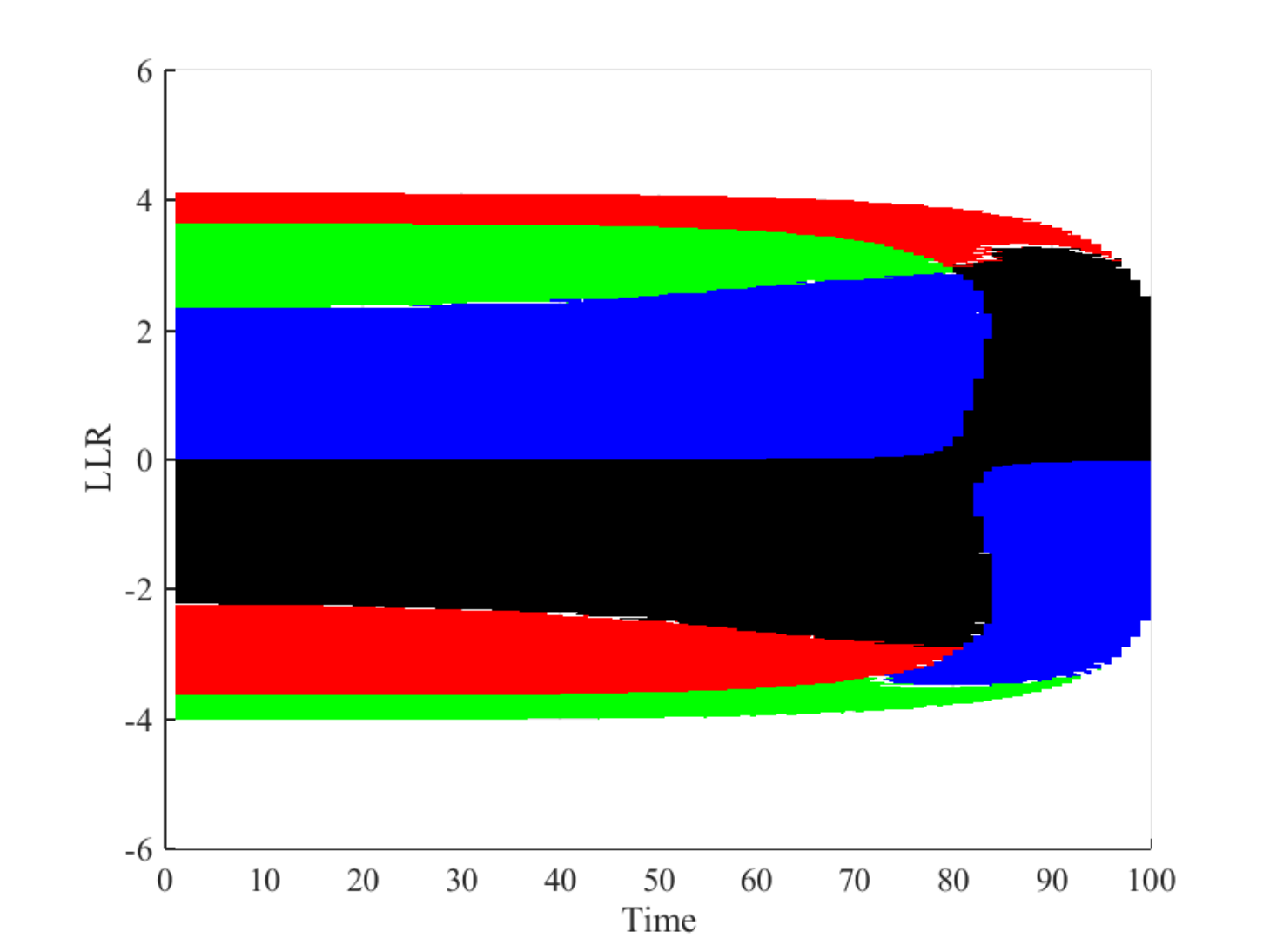}}
\subfigure[$T^1=6, T^2=9$]{\includegraphics[width=0.49\textwidth]{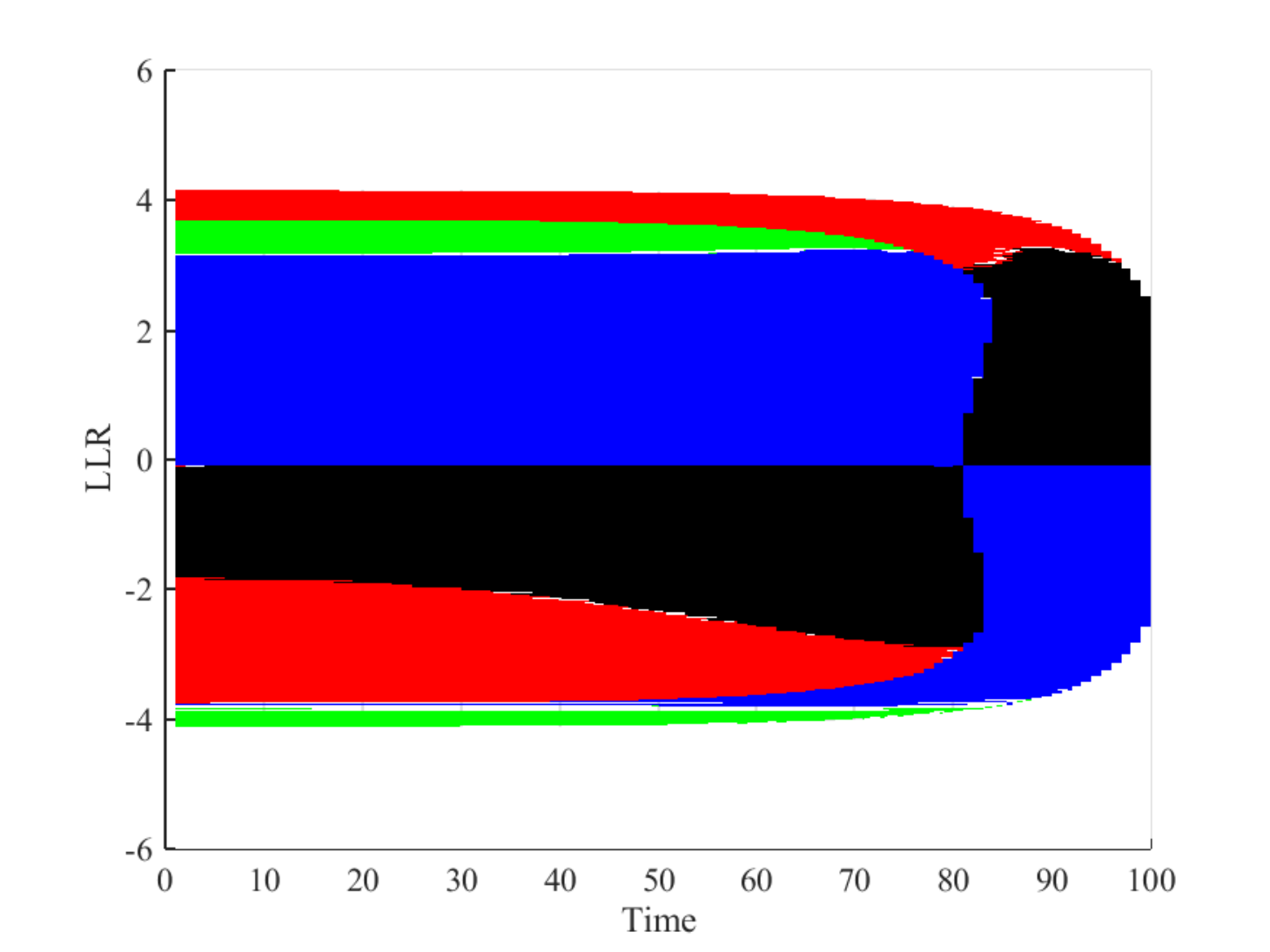}}
\caption{The stopping boundaries and selection region for $N=100$. We set $\alpha\approx0.01, \beta\approx 0.01$. Black: sensor 1. Blue: sensor 2. Red: sensor 3. Green: sensor 4.}
\label{fig:decision_region}
\end{figure}

Fig. \ref{fig:decision_region} illustrates the decision region of the $N$-horizon sequential test, including the stopping boundaries (i.e., $[-A_t, B_t]$) and selection function (i.e., $\psi_{t+1}(L_t)$). Note that, hereafter, we represent the results in terms of the sufficient statistic LLR, which is equivalent to the posterior given the prior. The  black, blue, red, and green colors represent the intervals within which Sensor 1, 2, 3, and 4 should be selected respectively. The following observations are made:
\begin{itemize}
\item The curved stopping boundaries comply with the result in Theorem 1-(b). 
\item The selection function $\psi_{t+1}(L_t)$ in Theorem 1-(a) is represented by simple partitions of the LLR domain. In specific, the fusion center decides the selected sensor at $t+1$ based on the region that $L_t$ resides in. Interestingly, the selection function from $t=1\to N$ is highly structured, and does not require large memory for storage.
\item The sensor usages are equal to the discrete time that LLR spends in the corresponding region before stopping. Thus the selection strategy controls the sensor usages by altering these selection regions. In Fig. \ref{fig:decision_region}-(a), if all sensors are constraint-free, then Sensor 1 and Sensor 2 are always preferred over the other two. Intuitively Sensor 1 dominates sensor 3, Sensor 2 dominates Sensor 4, since their KLDs under both hypotheses are larger. In Fig. \ref{fig:decision_region}-(b), if we impose the usage constraints on Sensors 1 and 2, then Sensors 3 and 4 are used more, thus the partition of LLR domain is reassigned to comply with the constraints. That is, the selection region for Sensor 1 is split mainly by Sensor 3, while that of Sensor 2 by Sensor 2. Fig. \ref{fig:decision_region}-(c) shows that the selection regions alter as the usage constraints change from $T^1=6, T^2=9$ to $T^1=7, T^2=7$. In specific, the selection region of Sensor 1 shrinks while that of Sensor 2 expands. 
\end{itemize}

\begin{figure}
\centering
\subfigure[]{\includegraphics[width=0.75\textwidth]{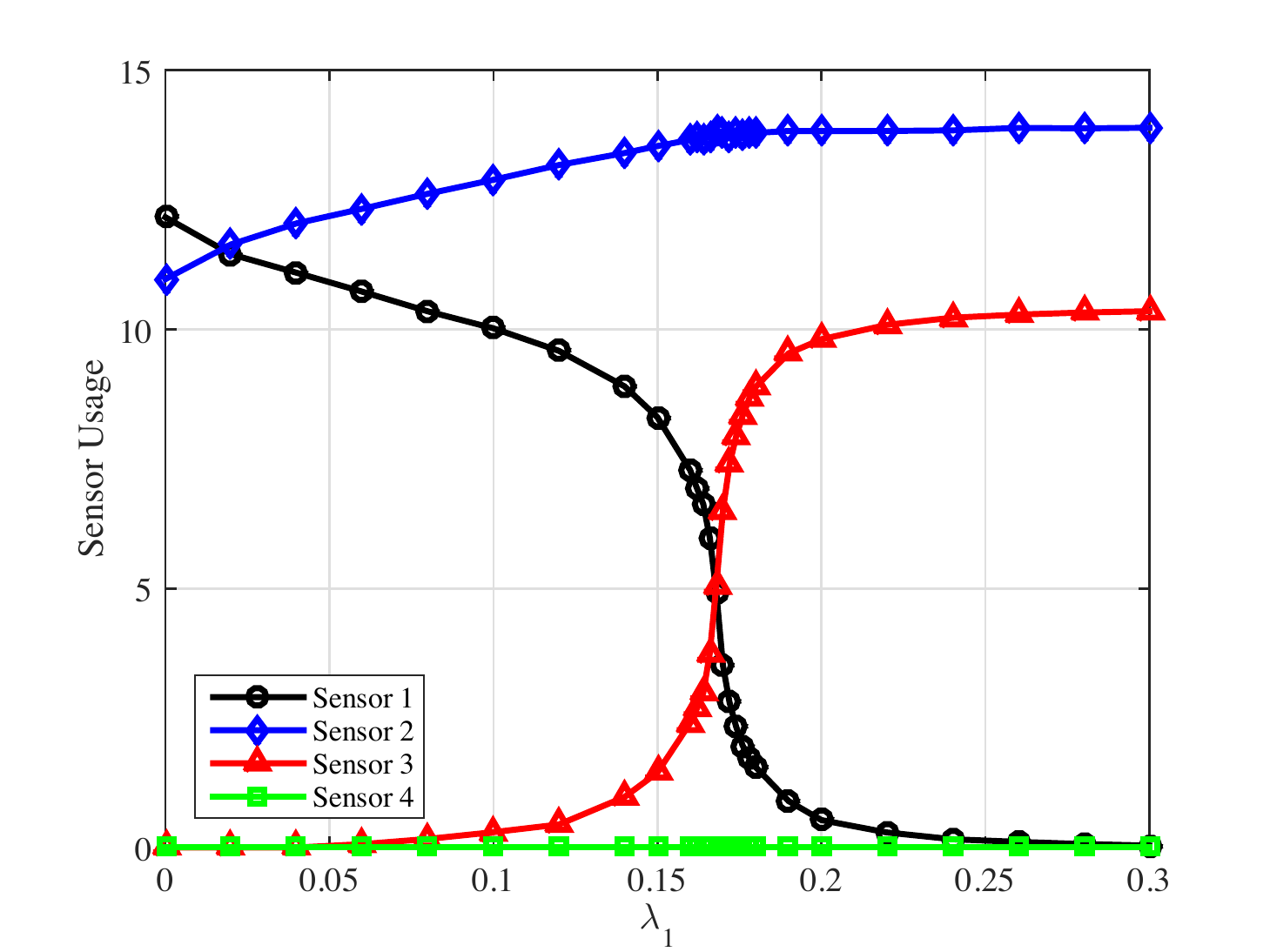}}
\subfigure[]{
\includegraphics[width=0.75\textwidth]{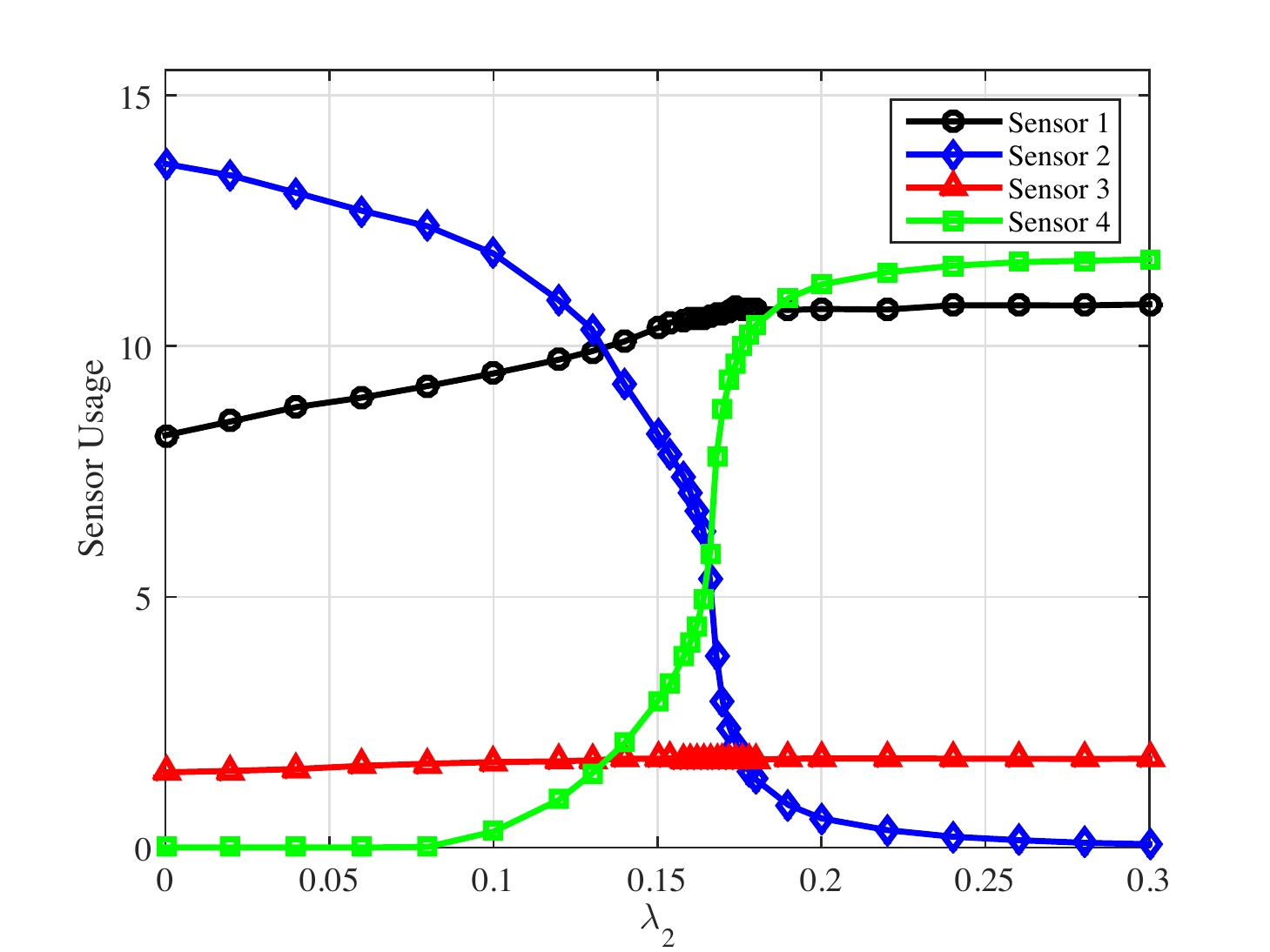}}
\caption{The sensor usage decreases as its associated multiplier increases. The error probabilities are set as $\alpha\approx0.0018, \beta\approx 0.0025$. (a) $\lambda_2=0$; (b) $\lambda_1=0.15$.}\label{fig:UsagevsLas}
\end{figure}

From Section III, we know that the selection regions, and thus the sensor usages, are governed by the multipliers, which are the parameters one can choose to meet the usage constraints. Bearing this in mind, Fig. \ref{fig:UsagevsLas} illustrates how we can control the sensor usages by setting the values of multipliers. 
In particular, Fig. \ref{fig:UsagevsLas}-(a) shows that the usage of Sensor 1 decreases from the full usage to zero as $\lambda_1$ increases, while other sensors increase their usages. Fig. \ref{fig:UsagevsLas}-(b) shows that the usage of Sensor 2 decreases to zero as $\la_2$ increases with fixed $\la_1=0.15$. 

\begin{figure}
\centering
\includegraphics[width=0.75\textwidth]{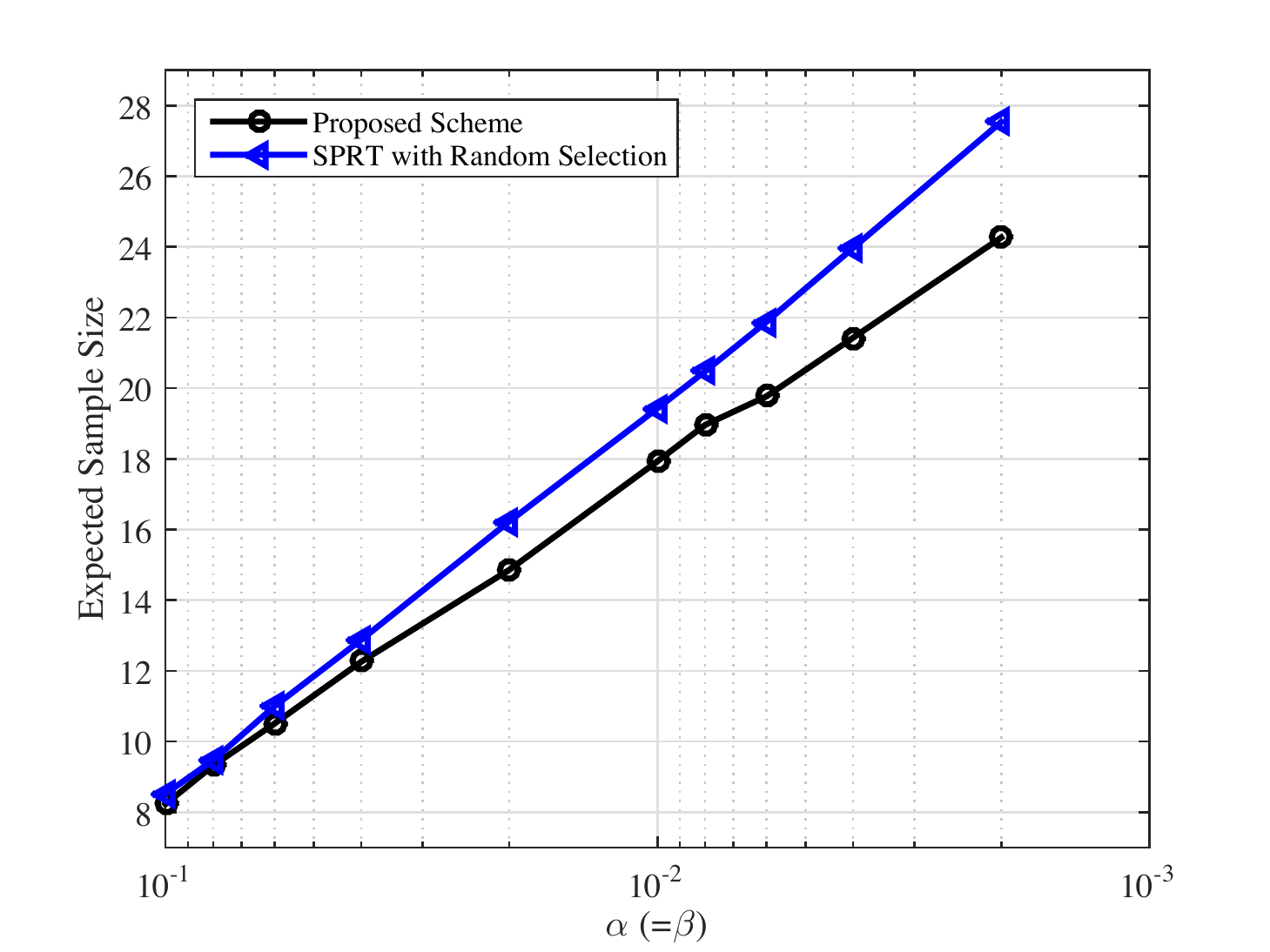}
\caption{Comparison of the proposed sequential test and the SPRT with offline random selection strategy.}\label{fig:ETvsErr}
\end{figure}

\begin{figure}
\centering
\includegraphics[width=0.75\textwidth]{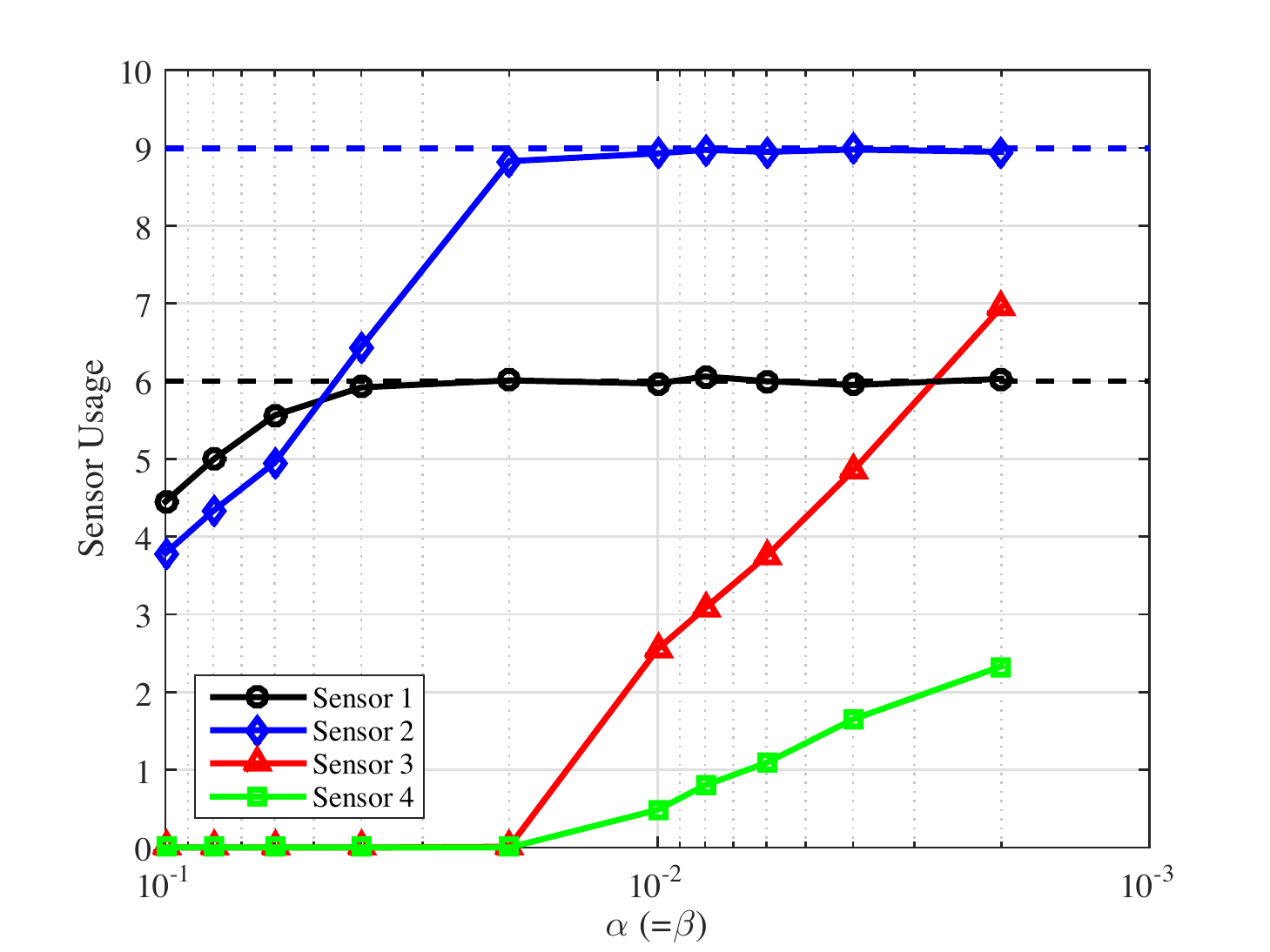}
\caption{Sensor usages of the proposed scheme corresponding to the experiment in Fig. \ref{fig:ETvsErr}.}\label{fig:UsagevsErr}
\end{figure}

Finally, in Fig. \ref{fig:ETvsErr}, we compare the proposed finite-$N$ sequential test with the existing method in \cite{Bai15}, which is an offline random selection algorithm. 
\ignore{The offline algorithm proposed in \cite{Bai15} build on solving a relaxation of the sequential hypothesis testing problem:
\begin{align*}
&\min \quad g({\bf q})\triangleq\pi_0 \frac{\alpha B-(1-\alpha)A}{{\bf D}^0{\bf q}}+\pi_1 \frac{(1-\beta) B-\beta A}{{\bf D}^1{\bf q}}\\
&\;\text{s.t.}\quad \,\; q_\ell\, g({\bf q})\le T^\ell, \;\;\ell\in \Omega_c
\end{align*}
One issue about this problem is that it is formulated from a relaxation of the original problem \eqref{P1}, and in practice, it is usually the case that error rates depend on both $A,B$ and the sensor selection scheme. }
The comparison is carried out at varying error probabilities $\alpha=\beta$, and fixed  sensor usage constraints for Sensor 1 and 2 ($T^1=6, T^2=9$, and Sensor 3 and 4 are free sensors). The corresponding multipliers are evaluated using the algorithm in Section IV. It is seen that the proposed online algorithm consistently outperforms the offline scheme with the same usage constraints and error probabilities. The improvement becomes more significant as the error probabilities decrease. Furthermore, Fig. \ref{fig:UsagevsErr} depicts the sensor usages of the proposed scheme in this experiment. When error probabilities are moderate ($\alpha=0.1 \to 0.06$ in Fig. \ref{fig:UsagevsErr}), Sensors 1 and 2 operate in free mode, and Sensors 3 and 4 are idle, which corresponds to the unconstrained scenario (i.e., the effective set of constraints are empty $\Omega_c=\emptyset$). This is similar to the case in Fig. \ref{fig:decision_region}-(a). As error rates decrease ($\alpha=0.04 \;\text{and}\;0.02$), Sensor 1 reaches the usage constraint first, while Sensor 2 still operates in free mode (i.e., $\Omega_c=\{1\}$). After $\alpha\le 0.01$, both Sensor 1 and 2 reach their usage limit and are under constraints (i.e., $\Omega_c=\{1, 2\}$). In this regime, we find multipliers such that constraints are satisfied with equalities. As error rates further decrease, free sensors like Sensors 3 and 4 are used more often, while Sensor 1 and 2 remain maximum usages at $T^1=6$ and $T^2=9$. 

\subsection{Infinite-Horizon}
In this subsection, the performance of the proposed scheme in the infinite-horizon setup is examined. We use a finite-horizon problem with sufficiently large $N=200$ to approximately evaluate the parameters (i.e., $A$, $B$ and selection regions) of the optimal sequential test. 

\begin{figure}
\centering
\subfigure[$T^1=6, T^2=9$]{\includegraphics[width=0.49\columnwidth]{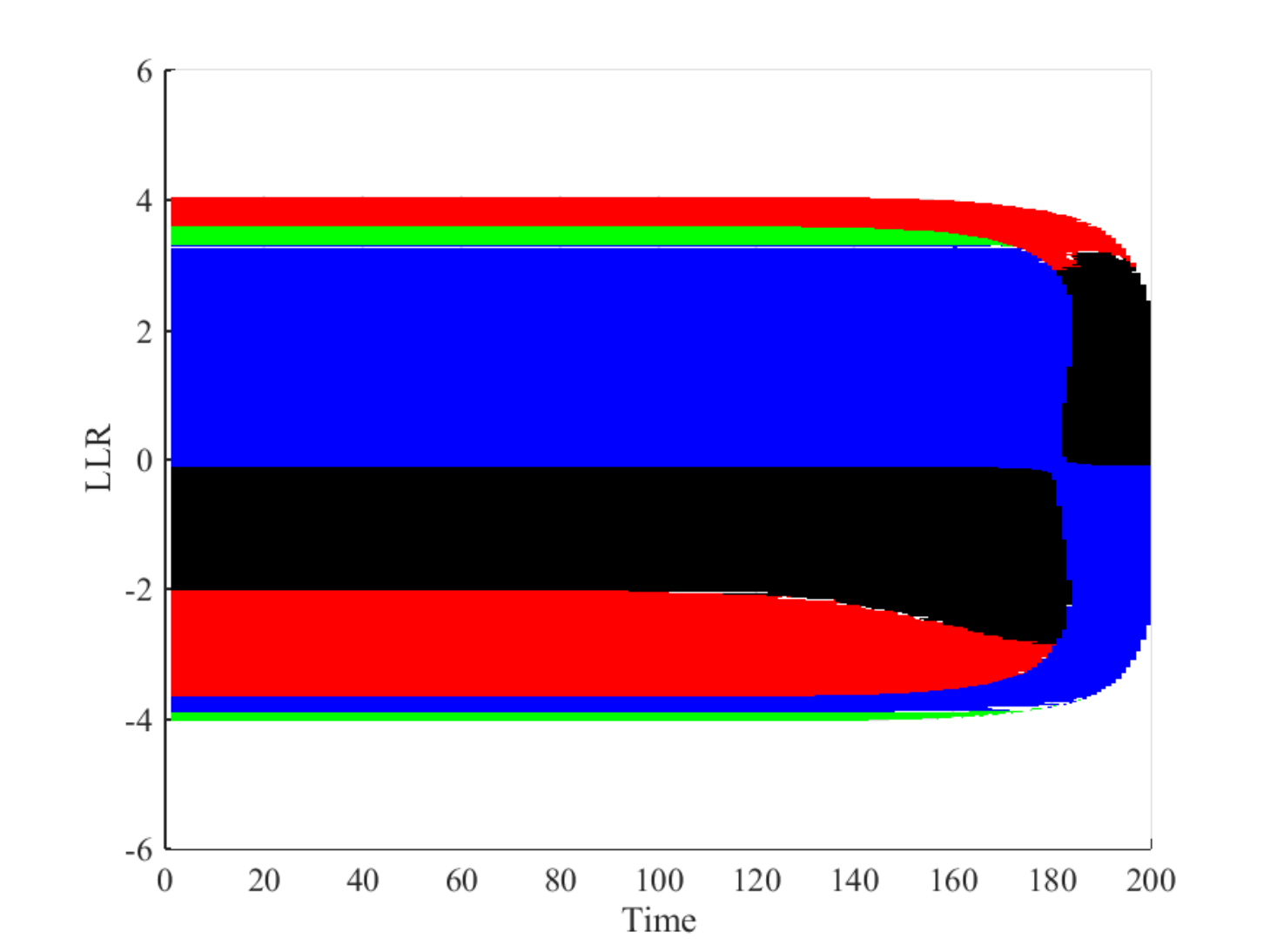}}
\subfigure[$T^1=7, T^2=7$]{\includegraphics[width=0.49\columnwidth]{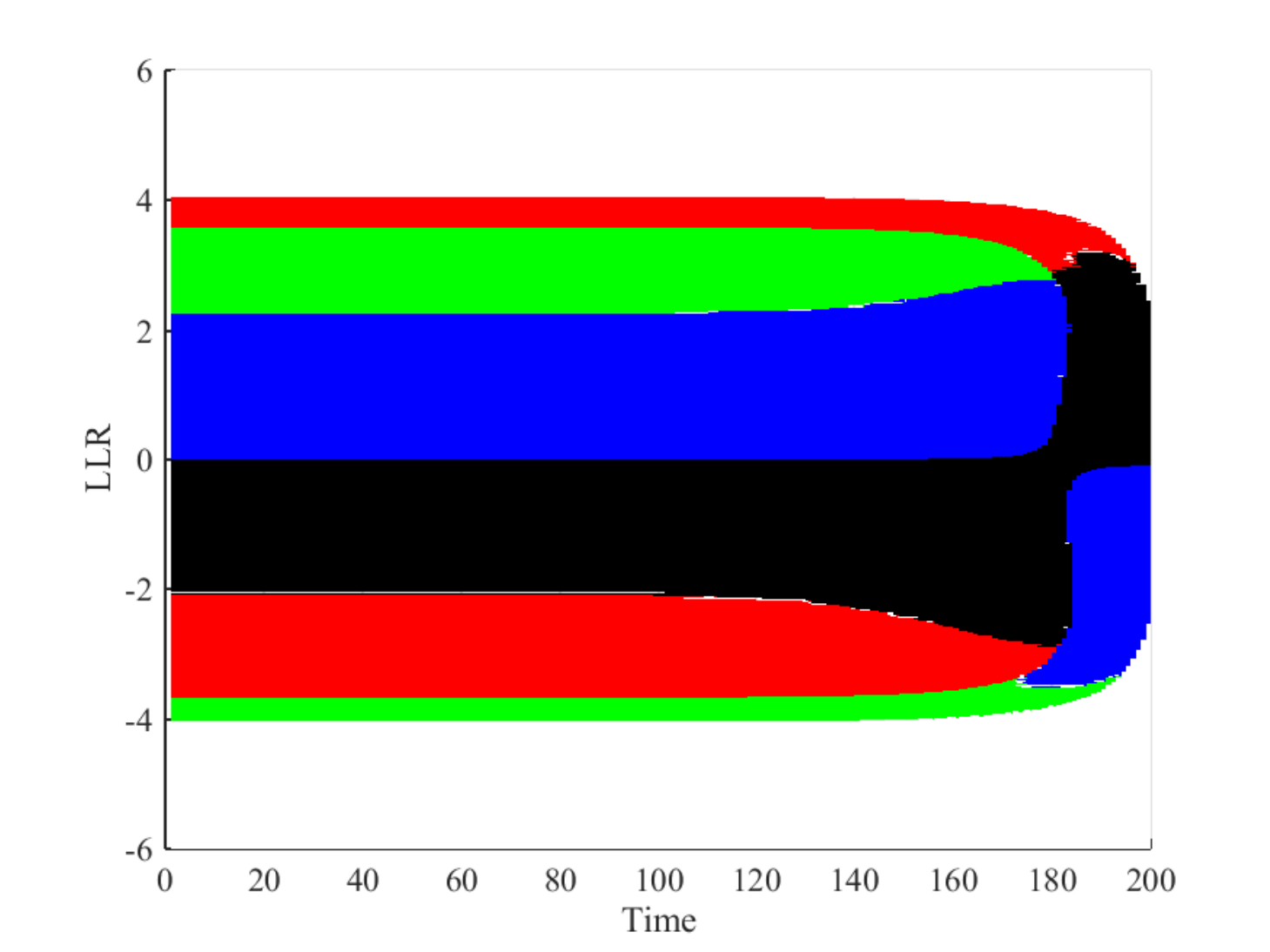}}
\caption{The stopping boundaries and selection function for $N=200$. We set $\alpha\approx0.01, \beta\approx 0.01$. Black: sensor 1. Blue: sensor 2. Red: sensor 3. Green: sensor 4.}\label{fig:Selection_region_200}
\end{figure}

\begin{figure}
\centering
\subfigure[$T^1=6, T^2=9$]{\includegraphics[width=0.49\columnwidth]{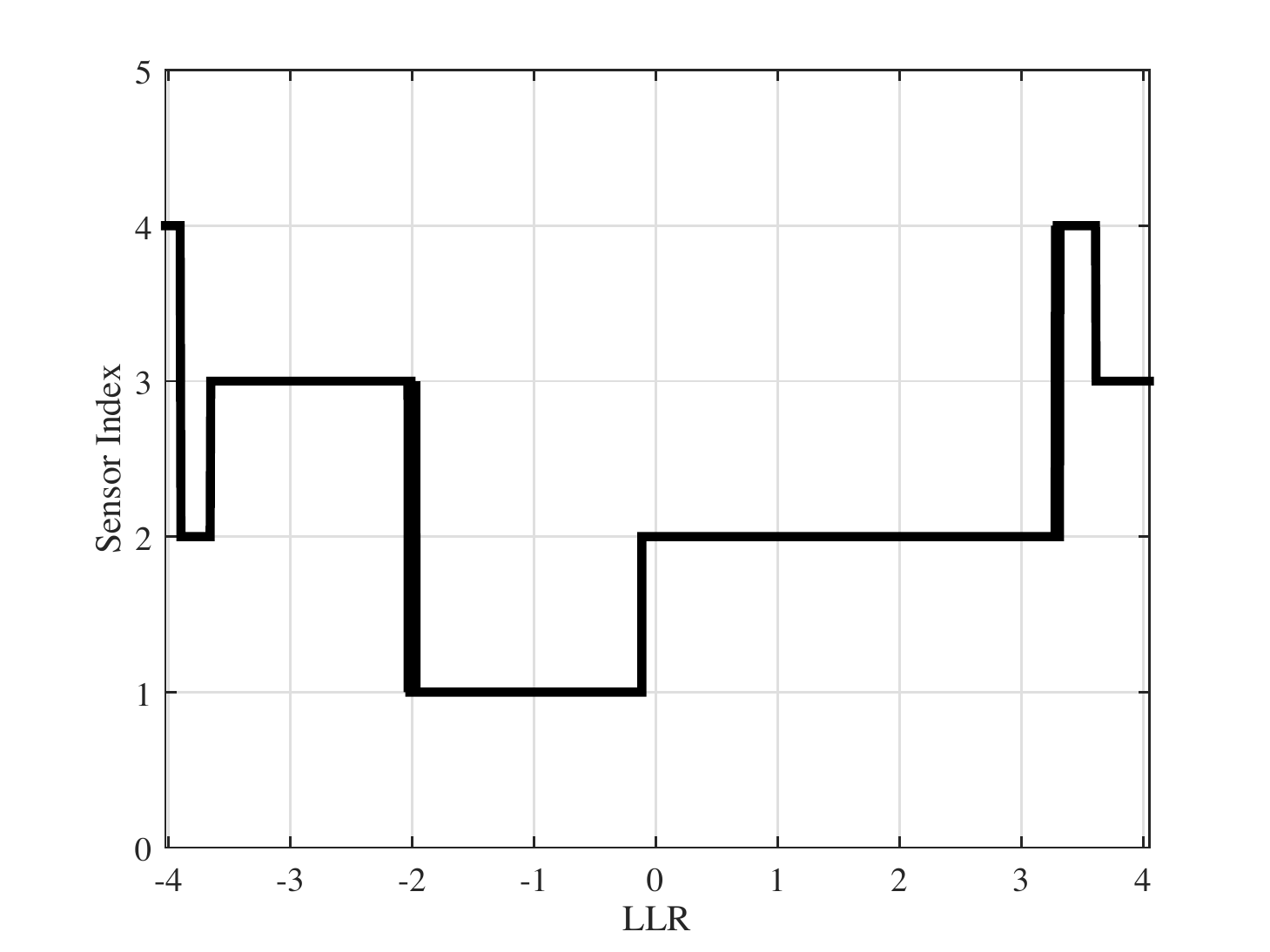}}
\subfigure[$T^1=7, T^2=7$]{\includegraphics[width=0.49\columnwidth]{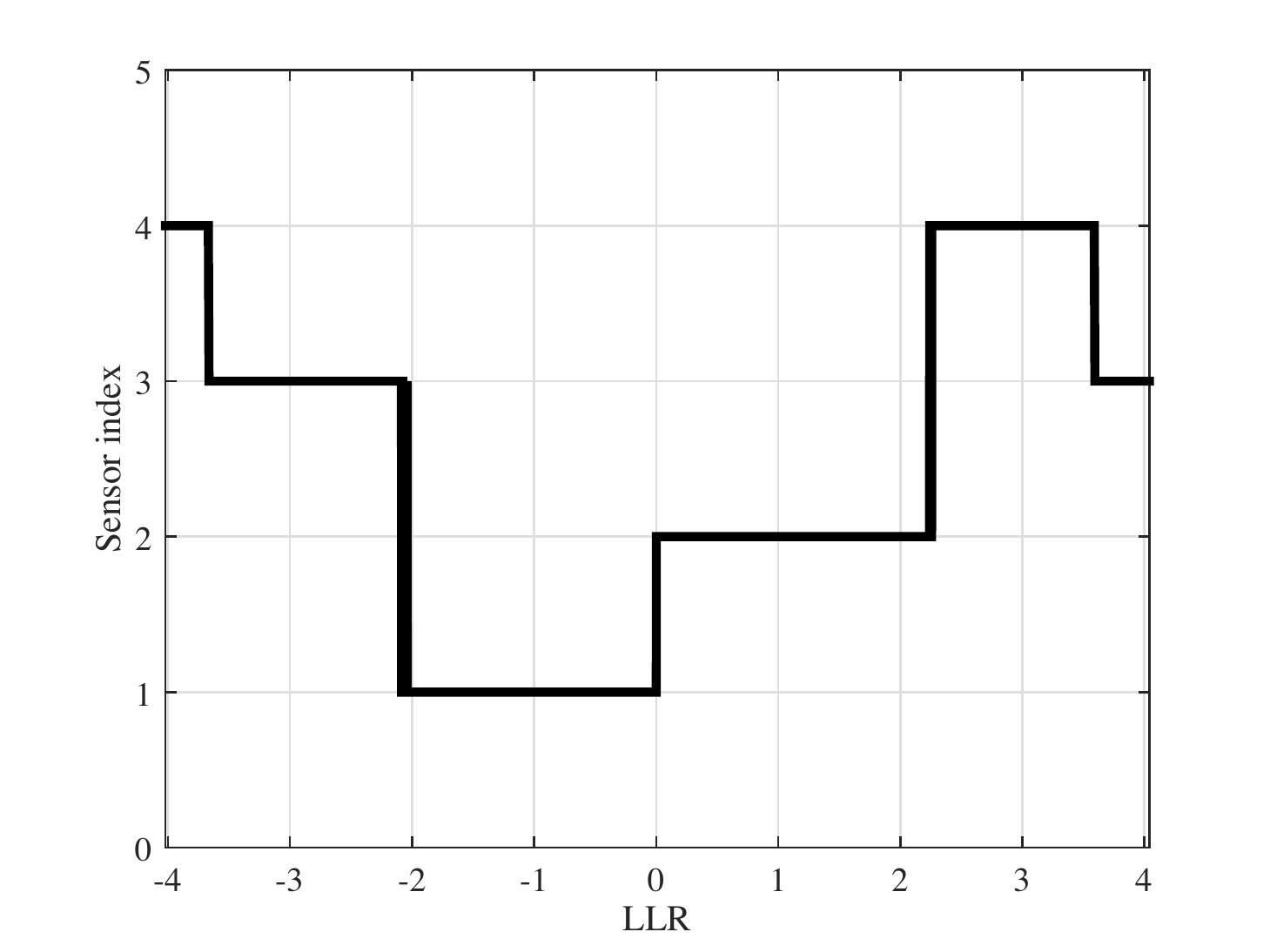}}
\caption{The stopping boundaries and selection intervals for the infinite-horizon problem. }\label{fig:Inf_SelectionIntervals}
\end{figure}

Again, Fig. \ref{fig:Selection_region_200} depicts the decision regions for the finite-horizon problem with $N=200$. Since a larger $N$ is used, compared to Fig. \ref{fig:decision_region}, Fig. \ref{fig:Selection_region_200} shows that the stopping boundaries and section strategy converge to the stable one at $t=0$, which is approximately the infinite-horizon solution according to Lemma 5. Unlike in the finite-horizon scenario, the fusion center only needs to store stopping boundaries and selection regions at $t=0$, which is depicted in Fig. \ref{fig:Inf_SelectionIntervals},  and use it for any $t$. This further lowers the storage demand. In specific, the selected sensor at $t+1$ is decided by which interval the LLR resides in at time $t$ within the stopping boundaries. We clearly see that the selection functions in Fig. \ref{fig:Inf_SelectionIntervals}-(a) change to that in Fig. \ref{fig:Inf_SelectionIntervals}-(b) as the usage constraints alter. 

\begin{figure}
\centering
\includegraphics[width=0.76\textwidth]{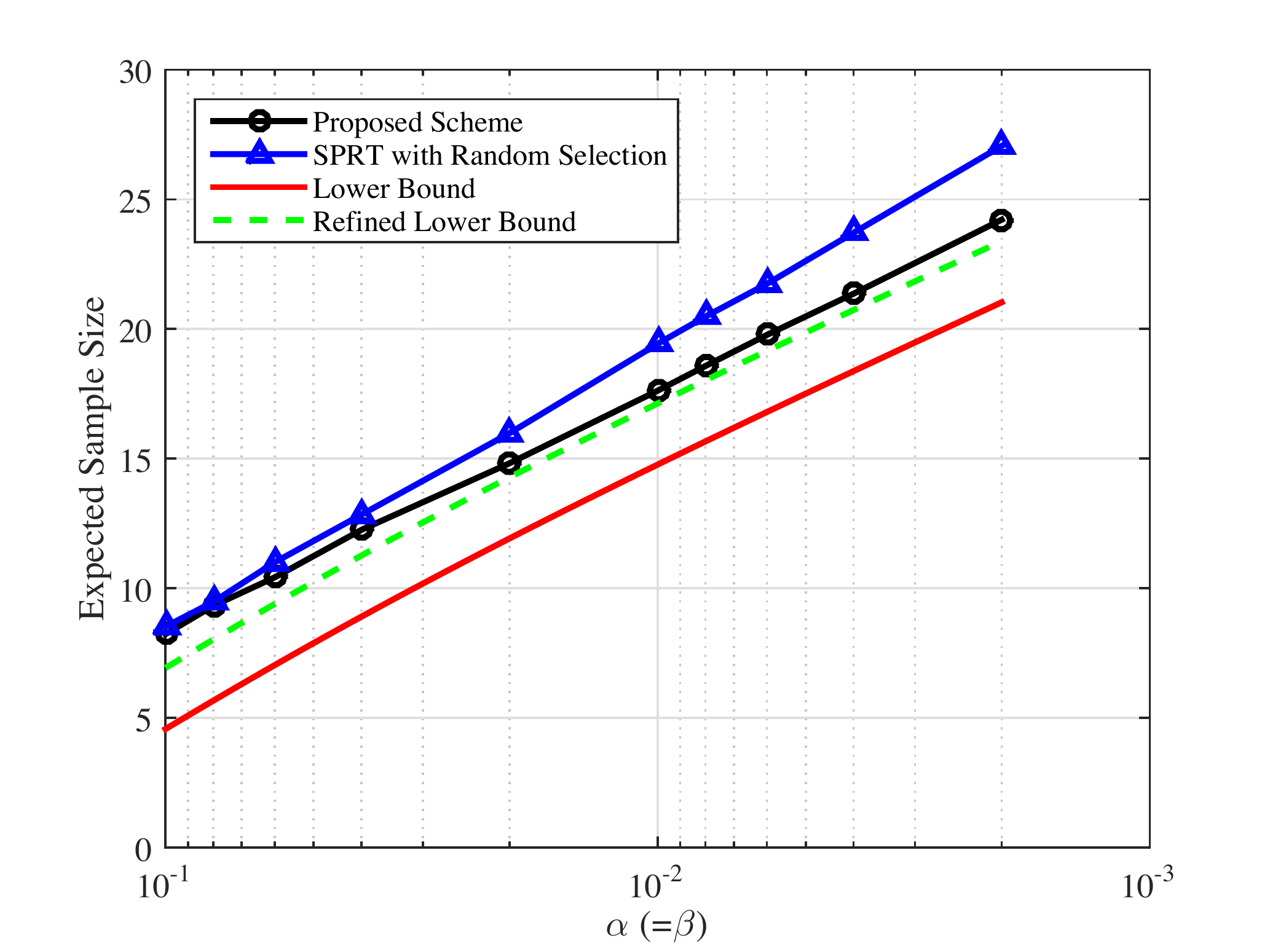}
\caption{Comparison of the proposed sequential test and the SPRT with offline random selection strategy.}\label{fig:comparison}
\end{figure}

Finally, in Fig. \ref{fig:comparison}, we compare the proposed scheme with the existing offline random selection scheme in \cite{Bai15}. Compared to Fig. \ref{fig:ETvsErr}, the expected sample size slightly decreases due to the removal of the hard limit on horizon $N$. Again, the proposed online scheme increasingly outperforms the offline selection scheme as the error probabilities become small. In addition, we also plot the close-form approximation for the optimal performance, which is given by \eqref{Bound}. Note that this analytical result (i.e., the red solid line) lies parallel to the performance curve of the proposed scheme (i.e., the black line with circle marks), indicating its accurate characterization for the asymptotical performance. The constant gap in between is largely caused by the inequality \eqref{bound_2} that lower bounds the constant term (i.e., independent of $\alpha$ and $\beta$) in \eqref{bound_1}, which ultimately leads to \eqref{Bound}. Therefore, the constant gap can be small if \eqref{bound_2} is tight, depending on the specific model. To see this, assuming that we derive the performance formula directly based on \eqref{bound_1} (specifically, $T_0^\ell$ and $T_1^\ell$ in \eqref{bound_1} need to be evaluated through simulation), it is shown in Fig. \ref{fig:comparison} that the resulting lower bound (i.e., the green dash line) aligns closely to the performance of the proposed scheme.  

\section{Conclusions}
In this work, we have studied the sequential hypothesis testing with online sensor selection and sensor usage constraints. The optimal sequential test and selection strategy are obtained for both the finite-horizon and infinite-horizon scenarios. We have also proposed algorithms to approximately evaluate the parameters in the optimal sequential procedure. Finally, extensive numerical results have been provided to illustrate the theoretical findings and comparison with the existing  method. Future works may include applying the same framework to address the usage-constrained sensor selection in other sequential problems, for example, change-point detection. Instead of the average sample size, other objective can also be studied, for example, the worst-case sample size. The applications in distributed sensor networks can be considered as well. For example, dynamic selection of quantization mode in the sequential detection \cite{Nguyen06,Mei08}.  
\section*{Appendix}
\proof[Proof of Lemma 1]
We want to prove that $g_n\lb X_{1:n}, \Bdelta_{1:n}\rb=g_n\lb \pi_1(n)\rb$.
It suffices to prove that for any realizations of $\{X_{1:n},\Bdelta_{1:n}\}$, i.e., $\{x_{1:n}, s_{1:n}\}$ and $\{\bar{x}_{1:{n}}, \bar{s}_{1:n}\}$, that lead to equal posteriors $\pi_1(n)=\bar{\pi}_1({{n}})$, we have $g_n\lb x_{1:n}, s_{1:n} \rb=g_{{n}}\lb \bar{x}_{1:n}, \bar{s}_{1:n}\rb$.

Conditioned on the event $\{\T=n\}$, by \eqref{stopping_rule}, it is obvious that $g_n\lb x_{1:n}, s_{1:n} \rb=g_{n}\lb \bar{x}_{1:n}, \bar{s}_{1:n}\rb=\phi\lb\pi_1(n)\rb$. Conditioned on the event $\{n<\T\le N\}$, we will prove by contradiction. On one hand, assume that $g_n\lb x_{1:n}, s_{1:n}\rb>g_{n}\lb \bar{x}_{1:n}, \bar{s}_{1:n}\rb$, then there exists a procedure $\left\{\tilde{\delta}_{n+1}, \{\tilde{\Bdelta}_{n+2:\widetilde{\T}}, \widetilde{\T}\}\in \MA_{n+1}^N\right\}$ (given $\{{x}_{1:n}, {s}_{1:n}\}$) such that
\begin{align}\label{contradiction_N}
g_n(x_{1:n},s_{1:n})&\ge \underbrace{\phi\lb \pi_1(n)\rb - \left[\E\lb \left.\phi\lb\pi_1(\widetilde{\T})\rb+\sum_{t=n+1}^{\widetilde{\T}} \MC_{\widetilde{\delta}_t}\right|X_{1:n}=x_{1:n}, \Bdelta_{1:n}=s_{1:n}\rb\right]}_{\widetilde{g}_n\lb x_{1:n}, s_{1:n}\rb}\nonumber\\&>{g_n\lb \bar{x}_{1:n}, \bar{s}_{1:n}\rb},
\end{align}
due to the definition of $g_n$ in \eqref{stopping_rule}.


On the other hand, we construct the following procedure $\left\{\widehat{\delta}_{n+1}, \{\widehat{\Bdelta}_{n+2:{\widehat{\T}}}, \widehat{\T}\}\in \MA^N_{n+1}\right\}$  (given $\{\bar{x}_{1:n}, \bar{s}_{1:n}\}$). Let 
\begin{align}\label{construction_1}
\widehat \delta_{n+1}(\bar{x}_1, \ldots, \bar{x}_n)=\widetilde \delta_{n+1}({x}_1, \ldots, {x}_n),
\end{align} 
and, given the same samples after time $n$ (denoted as $x_{n+1}, x_{n+2}, \ldots$), 
\begin{align}\label{construction_2}
\widehat \delta_{t}(\bar{x}_1, \ldots, \bar{x}_n, x_{n+1}, \ldots, x_{t-1})=\widetilde \delta_{t}({x}_1, \ldots, {x}_n, x_{n+1}, \ldots, x_{t-1}), \quad t=n+2, \ldots, N. 
\end{align}
Moreover,  let $\widehat{\T}$ stop  if $\widetilde{\T}$ stops given the same samples $\{x_{n+1}, x_{n+2},\ldots\}$, and the decision rule $$\widehat D(\bar{x}_1, \ldots, \bar{x}_n, x_{n+1}, \ldots, x_{\widehat\T})=\widetilde D(\bar{x}_1, \ldots, \bar{x}_n, x_{n+1}, \ldots, x_{\widetilde\T}).$$ { In short,  the procedure $\left\{\widehat{\Bdelta}_{n+1:{\widehat{\T}}}, \widehat{\T}\right\}$ is designed to yield the exact same actions as that of the procedure $\left\{\widetilde{\Bdelta}_{n+1:{\widetilde{\T}}}, \widetilde{\T}\right\}$ given the same samples at time $n$, i.e.,  $\{x_{n+1}, x_{n+2}, \ldots\}$.
Note that, according to the above construction process, $\left\{\widehat{\Bdelta}_{n+1:{\widehat{\T}}}, \widehat{\T}\right\}$ and $\left\{\widetilde{\Bdelta}_{n+1:{\widetilde{\T}}}, \widetilde{\T}\right\}$ are not identical procedures since $\{{x}_{1:n}, {s}_{1:n}\}\neq \{\bar{x}_{1:n}, \bar{s}_{1:n}\}$.}

Again, due to the definition of $g_n$ in \eqref{stopping_rule}, we also have
\begin{align}\label{proof_pre0}
g_n\lb \bar{x}_{1:n}, \bar{s}_{1:n}\rb\ge
\underbrace{\phi\lb \pi_1(n)\rb - \left[\E\lb \left.\phi\lb\pi_1(\hat{\T})\rb+\sum_{t=n+1}^{\hat{\T}} \MC_{\hat{\delta}_t}\right|{X}_{1:n}=\bar{x}_{1:n}, {\Bdelta}_{1:n}=\bar{s}_{1:n}\rb\right]}_{\widehat{g}_n\lb \bar{x}_{1:n}, \bar{s}_{1:n}\rb}.
\end{align}

Next, we prove that 
\begin{align}\label{proof_0}
\widehat{g}\lb\bar{x}_{1:n}, \bar{s}_{1:n}\rb=\widetilde{g}\lb{x}_{1:n}, {s}_{1:n}\rb,
\end{align}
which requires 
\begin{align}\label{proof_1}
\E\!\!\lb\!\! \left.\phi\lb\pi_1(\widetilde{\T})\rb\!\!+\!\!\sum_{t=n+1}^{\widetilde{\T}} \MC_{\widetilde{\delta}_t}\right|{X}_{1:n}\!=\!{x}_{1:n}, {\Bdelta}_{1:n}\!=\!{s}_{1:n}\rb\!=\!\E\!\!\lb\!\! \left.\phi\lb\pi_1(\widehat{\T})\rb\!\!+\!\!\sum_{t=n+1}^{\widehat{\T}}\MC_{\widehat{\delta}_t}\right|{X}_{1:n}\!=\!\bar{x}_{1:n}, {\Bdelta}_{1:n}\!=\!\bar{s}_{1:n}\rb.
\end{align}
First, due to the construction of $\left\{\widehat{\Bdelta}_{n+1:{\widehat{\T}}}, \widehat{\T}\right\}$, we have 
\begin{align}\label{proof_2}
\left.\widetilde{\T}-n\right|\{X_{1:n}=\bar{x}_{1:n}, \Bdelta_{1:n}=\bar{s}_{1:n}\}=\left.\widehat{\T}-n\right|\{X_{1:n}={x}_{1:n}, \Bdelta_{1:n}={s}_{1:n}\},\quad \text{a.s.}.
\end{align}
To show that the first terms on both sides of \eqref{proof_1} are equal, i.e., 
\begin{align}
\E\lb \left.\phi\lb\pi_1(\widetilde{\T})\rb\right|X_{1:n}, \Bdelta_{1:n}\rb=\E\lb \left.\phi\lb\pi_1(\bar{\T})\rb\right|\bar{X}_{1:n}, \bar{\Bdelta}_{1:n}\rb,
\end{align}
notice that
\begin{align}
\pi_1(\widetilde{\T})=\frac{\pi_1(n)e^{\sum_{t=n+1}^{\widetilde{\T}} l_{\widetilde{\delta}_t}}}{\pi_0(n)+\pi_1(n)e^{\sum_{t=n+1}^{\widetilde{\T}} l_{\widetilde{\delta}_t}}}
\end{align}
has the same distribution conditioned on $\{\bar{x}_{1:n}, \bar{s}_{1:n}\}$ as that of 
\begin{align}
\pi_1({\widehat{\T}})=\frac{\bar{\pi}_1(n)e^{{\sum_{t=n+1}^{{\widehat\T}} l_{\widehat{\delta}_t}}}}{\bar{\pi}_0(n)+\bar{\pi}_1(n)e^{\sum_{t=n+1}^{\widehat{\T}} l_{\widehat{\delta}_t}}}
\end{align}
conditioned on  $\{{x}_{1:n}, {s}_{1:n}\}$. This is true because $\pi_1(n)=\bar{\pi}_1(n)$ and $\sum_{t=n+1}^{\widetilde\T} l_{\widetilde\delta_t}$ has the same posterior distribution as $\sum_{t=n+1}^{\widehat{\T}} l_{\widehat{\delta}_t}$ due to \eqref{construction_1}-\eqref{construction_2} and \eqref{proof_2}. In addition, the second terms on both sides of \eqref{proof_1} are also equal by combining  \eqref{construction_1}-\eqref{construction_2} and \eqref{proof_2}.

Using \eqref{proof_pre0}-\eqref{proof_0}, we arrive at
\begin{align}
g_{n}\lb \bar{x}_{1:n}, \bar{s}_{1:n}\rb\ge \widehat{g}_{n}\lb\bar{x}_{1:n}, \bar{s}_{1:n}\rb=\widetilde{g}_n\lb x_{1:n}, s_{1:n}\rb
\end{align}
which contradicts with \eqref{contradiction_N}. 

Similar contradiction appears if we assume $g_n\lb X_{1:n}, \Bdelta_{1:n}\rb<g_{n}\lb \bar{X}_{1:n}, \bar{\Bdelta}_{1:n}\rb$.
\endproof
\proof[Proof of Lemma 4]
By the concavity of $\widetilde{\MG}_n^N(x)$, we know that the continuation region at $t=n$ is an interval confined by the roots of the following equations (denoted as $a_n$ and $b_n$ respectively):
\begin{align}
\mu_0(1-x)\widetilde{\MG}_n^N(x),  \;\text{and}\;\;\mu_1x=\widetilde{\MG}_n^N(x), \quad n<N.
\end{align}
Since $\widetilde{\MG}_{n-1}^N(x)<\widetilde{\MG}_n^N(x)$, thus $a_{n-1}<a_{n}$ and $b_{n-1}>b_{n}$. At $t=N$, the procedure has to stop and make decision, thus $a_N=b_N$. $\mu_1\pi_1(N)\,\substack{>\\<}\,\mu_0\pi_0(N)$ which gives  $\pi_1(N)\,\substack{>\\<}\,a_N=\mu_0/(\mu_0+\mu_1)$.
\endproof

\proof[Proof of Proposition \ref{cor1}]
Note that for the LLR statistic, we have
\begin{align}\label{Bound_0}
\E_0\lb L_\T\rb=\E_0\left[\sum_{t=1}^\T\lb\sum_{\ell\in\Omega_c}l_{\delta_t}\mathbbm{1}_{\{\delta_t=\ell\}}+l_{\delta_t}\mathbbm{1}_{\{\delta_t\in \overline{\Omega}_c\}}\rb\right].
\end{align}
The first term of \eqref{Bound_0} can be expressed as
\begin{align}\label{Bound_0_term1}
\E_0\left[\sum_{t=1}^\T\sum_{\ell\in\Omega_c}l_{\delta_t}\mathbbm{1}_{\{\delta_t=\ell\}}\right]=&\sum_{\ell\in\Omega_c}\E_0\lb \sum_{t=1}^\infty l_{\delta_t}\mathbbm{1}_{\{\delta_t=\ell\}}\mathbbm{1}_{\{\T\ge t\}}\rb\nonumber\\=&\sum_{\ell\in\Omega_c}\E_0\lb\sum_{t=1}^\infty\E_0\lb\left. l_{\delta_t}\right|X_{1:(t-1)}, \Bdelta_{1:t-1}\rb\mathbbm{1}_{\{\delta_t=\ell\}}\mathbbm{1}_{\{\T\ge t\}}\rb\nonumber\\=&-\sum_{\ell\in\Omega_c}D^\ell_0\,\E_0\lb\sum_{t=1}^\infty\mathbbm{1}_{\{\delta_t=\ell\}}\mathbbm{1}_{\{\T\ge t\}}\rb\nonumber\\=&-\sum_{\ell\in\Omega_c} D_0^\ell T_0^\ell,
\end{align}
where $D_0^\ell\triangleq \E_0\lb -l_\ell \rb$ is the KL divergence of sensor $\ell$ and $T^\ell_0\triangleq \E_0\lb\sum_{t=1}^\T \mathbbm{1}_{\{\delta_t=\ell\}}\rb$ is the mean usage under $\MH_0$. Furthermore, the second term of \eqref{Bound_0} can be bounded as follows
\begin{align}
\E_0\left[\sum_{t=1}^\T l_{\delta_t}\mathbbm{1}_{\{\delta_t\in \overline{\Omega}_c\}}\right]=&\E_0\lb\sum_{t=1}^\infty l_{\delta_t}\mathbbm{1}_{\{\delta_t\in \overline{\Omega}_c\}}\mathbbm{1}_{\{\T\ge t\}}\rb\nonumber\\=&\E_0\lb\sum_{t=1}^\infty \E_0\lb\left. l_{\delta_t}\right|X_{1:(t-1)}, \Bdelta_{1:t-1}\rb\mathbbm{1}_{\{\delta_t\in \overline{\Omega}_c\}}\mathbbm{1}_{\{\T\ge t\}}\rb\label{Bound_0_ineq}\\\ge&- \max_{\ell\in\overline{\Omega}_c} D_0^\ell \,\E_0\lb\sum_{t=1}^\infty  \mathbbm{1}_{\{\delta_t\in \overline{\Omega}_c\}}\mathbbm{1}_{\{\T\ge t\}}\rb\nonumber\\=&-\max_{\ell\in\overline{\Omega}_c} D_0^\ell \lb\E_0\T-\sum_{\ell\in\Omega_c}T_0^\ell \rb,\label{Bound_0_term2}
\end{align}
where inequality \eqref{Bound_0_ineq} holds because $\E_0\lb\left. l_{\delta_t}\right|X_{1:(t-1)}, \Bdelta_{1:t-1}\rb\mathbbm{1}_{\{\delta_t\in \overline{\Omega}_c\}}\ge \min_{\ell\in\overline{\Omega}_c} \E_0\lb l_\ell\rb\mathbbm{1}_{\{\delta_t\in \overline{\Omega}_c\}}$.
Applying \eqref{Bound_0_term1} and \eqref{Bound_0_term2} to \eqref{Bound_0} results in
\begin{align}
\E_0\lb L_\T\rb\ge -\sum_{\ell\in \Omega_c}D_0^\ell T_0^\ell -\max_{\ell\in\overline{\Omega}_c} D_0^\ell \lb\E_0\T-\sum_{\ell\in\Omega_c}T_0^\ell\rb,
\end{align}
which leads to the bound for mean sample size under $\MH_0$:
\begin{align}
\E_0\T&\ge \left[-\E_0\lb L_\T\rb-\sum_{\ell\in \Omega_c}D_0^\ell T_0^\ell +\max_{\ell\in\overline{\Omega}_c} D_0^\ell\sum_{\ell\in\Omega_c}T_0^\ell\right]\frac{1}{\max_{\ell\in\overline{\Omega}_c} D_0^\ell}\nonumber\\&=\frac{-\E_0\lb L_\T\rb}{\max_{\ell\in\overline{\Omega}_c} D_0^\ell}+\sum_{\ell\in\Omega_c}\lb 1-\frac{D^\ell_0}{\max_{\ell\in\overline{\Omega}_c} D_0^\ell}\rb T^\ell_0.\label{Bound_H0}
\end{align}
Under $\MH_1$, similarly as in \eqref{Bound_0_term1} and \eqref{Bound_0_term2}, we have
\begin{align}
\E_1\left[\sum_{t=1}^\T\sum_{\ell\in\Omega_c}l_{\delta_t}\mathbbm{1}_{\{\delta_t=\ell\}}\right]&=\sum_{\ell\in\Omega_c} D_1^\ell T_1^\ell,\\
\text{and}\quad \E_1\left[\sum_{t=1}^\T l_{\delta_t}\mathbbm{1}_{\{\delta_t\in \overline{\Omega}_c\}}\right]&=\E_1\lb\sum_{t=1}^\infty l_{\delta_t}\mathbbm{1}_{\{\delta_t\in \overline{\Omega}_c\}}\mathbbm{1}_{\{\T\ge t\}}\rb
\nonumber\\&=\E_1\lb\sum_{t=1}^\infty \E_1\lb l_{\delta_t}|\mathcal{F}_{t-1}\rb\mathbbm{1}_{\{\delta_t\in \overline{\Omega}_c\}}\mathbbm{1}_{\{\T\ge t\}}\rb
\nonumber\\&\le \max_{\ell\in\overline{\Omega}_c} D_1^\ell \,\E_1\lb\sum_{t=1}^\infty  \mathbbm{1}_{\{\delta_t\in \overline{\Omega}_c\}}\mathbbm{1}_{\{\T\ge t\}}\rb
\nonumber\\&=\max_{\ell\in\overline{\Omega}_c} D_1^\ell \lb\E_1\T-\sum_{\ell\in\Omega_c}T_1^\ell \rb,
\end{align}
that lead to
\begin{align}
\E_1\lb L_\T\rb&=\E_1\left[\sum_{t=1}^\T\sum_{\ell\in\Omega_c}l_{\delta_t}\mathbbm{1}_{\{\delta_t=\ell\}}\right]+\E_1\left[\sum_{t=1}^\T l_{\delta_t}\mathbbm{1}_{\{\delta_t\in \overline{\Omega}_c\}}\right]\nonumber\\&\le \sum_{\ell\in \Omega_c}D_1^\ell T_1^\ell +\max_{\ell\in\overline{\Omega}_c} D_1^\ell \lb\E_1\T-\sum_{\ell\in\Omega_c}T_1^\ell\rb .
\end{align}
As a result, we can bound the mean sample size under $\MH_1$ by
\begin{align}
\E_1\T&\ge \left[\E_1\lb L_\T\rb-\sum_{\ell\in \Omega_c}D_1^\ell T_1^\ell +\max_{\ell\in\overline{\Omega}_c} D_1^\ell\sum_{\ell\in\Omega_c}T_1^\ell\right]\frac{1}{\max_{\ell\in\overline{\Omega}_c} D_1^\ell}\nonumber\\&=\frac{\E_1\lb L_\T\rb}{\max_{\ell\in\overline{\Omega}_c} D_1^\ell}+\sum_{\ell\in\Omega_c}\lb 1-\frac{D^\ell_1}{\max_{\ell\in\overline{\Omega}_c} D_1^\ell}\rb T^\ell_1.\label{Bound_H1}
\end{align}
Finally, the expected mean sample size, i.e., $\E\T=\pi_0\E_0\T+\pi_1\E_1\T$, can be bounded below as follows:
\begin{align}
\E\T&\ge \pi_0\frac{-\E_0\lb L_\T\rb}{\max_{\ell\in\overline{\Omega}_c} D_0^\ell}+\pi_1\frac{\E_1\lb L_\T\rb}{\max_{\ell\in\overline{\Omega}_c} D_1^\ell}+\sum_{\ell\in\Omega_c}T^\ell-\sum_{\ell\in\Omega_c}\lb \frac{\pi_0D^\ell_0}{\max_{\ell\in\overline{\Omega}_c} D_0^\ell}T^\ell_0+\frac{\pi_1D^\ell_1}{\max_{\ell\in\overline{\Omega}_c} D_1^\ell}T^\ell_1\rb\label{bound_1}\\&\ge\pi_0\frac{-\E_0\lb L_\T\rb}{\max_{\ell\in\overline{\Omega}_c} D_0^\ell}+\pi_1\frac{\E_1\lb L_\T\rb}{\max_{\ell\in\overline{\Omega}_c} D_1^\ell}+\sum_{\ell\in\Omega_c}\lb1-\max\left\{\frac{D^\ell_1}{\max_{\ell\in\overline{\Omega}_c} D_1^\ell},\frac{D^\ell_0}{\max_{\ell\in\overline{\Omega}_c} D_0^\ell}\right\}\rb T^\ell,\label{bound_2}
\end{align}
where the second inequality is obtained by noting that $\pi_0T^\ell_0+\pi_1T^\ell_1=T^\ell$, thus 
\begin{align}\label{bound_inq}
\frac{\pi_0D^\ell_0}{\max_{\ell\in\overline{\Omega}_c} D_0^\ell}T^\ell_0+\frac{\pi_1D^\ell_1}{\max_{\ell\in\overline{\Omega}_c} D_1^\ell}T^\ell_1\le\max\left\{\frac{D^\ell_1}{\max_{\ell\in\overline{\Omega}_c} D_1^\ell},\frac{D^\ell_0}{\max_{\ell\in\overline{\Omega}_c} D_0^\ell}\right\}T^\ell,
\end{align}
with equality holds if $T^\ell=\pi_iT_i^\ell, \; i=\arg \max \left\{\frac{D^\ell_1}{\max_{\ell\in\overline{\Omega}_c} D_1^\ell},\frac{D^\ell_0}{\max_{\ell\in\overline{\Omega}_c} D_0^\ell}\right\}$.

Next, by drawing on the Wald's approximation \cite{SeqA_book}, i.e., $L_{\T^\star}\approx -A$ given $D^\star_{\T^\star}=0$ or $L_{\T^\star}\approx B$ given $D^\star_{\T^\star}=1$, we obtain
\begin{align}
\E_0\lb L_{\T^\star}\rb&=\alpha \E_0\lb L_{\T^\star}|D^\star_{\T^\star}=1\rb +(1- \alpha) \E_0\lb L_{\T^\star}|D^\star_{\T^\star}=0\rb \nonumber\\&=\alpha B-(1-\alpha) A,\label{appendix1}\\
\E_1\lb L_{\T^\star}\rb&=(1-\beta) \E_1\lb L_{\T^\star}|D^\star_{\T^\star}=1\rb +\beta \E_1\lb L_{\T^\star}|D^\star_{\T^\star}=0\rb \nonumber\\&=(1-\beta)B-\beta A.\label{appendix2}
\end{align}
Moreover,  invoking the change of measure technique and the Wald's approximation, we have
\begin{align}
&\alpha=\E_0\lb\mathbbm{1}_{\{D^\star_{\T^\star}=1\}}\rb=\E_1\lb\mathbbm{1}_{\{D^\star_{\T^\star}=1\}}e^{-L_{\T^\star}}\rb\approx e^{-B}\lb 1-\beta\rb,\\&\beta=\E_1\lb\mathbbm{1}_{\{D^\star_{\T^\star}=0\}}\rb=\E_0\lb\mathbbm{1}_{\{D^\star_{\T^\star}=0\}}e^{L_{\T^\star}}\rb\approx e^{-A}\lb 1-\alpha\rb,
\end{align}
which lead to
\begin{align}\label{appendix3}
B\approx \log \frac{1-\beta}{\alpha}, \quad A\approx \log \frac{1-\alpha}{\beta}.
\end{align} 
Substituting \eqref{appendix3} into \eqref{appendix1}-\eqref{appendix2} gives
$\E_0(L_{\T^\star})\approx -\MD(\alpha||1-\beta)$ and $\E_1(L_{\T^\star})\approx \MD(1-\beta||\alpha)$. 
\endproof

\ignore{\begin{table*}
\centering
\caption{{\color{red} This is table is for online experiment recording Fig. 2}}
  \begin{tabular}{| c | c | c | c | c | c | c | c | c |}
    \hline
 Fig. & $\alpha=\beta$ & $\{\mu_0, \mu_1\}$ &$\{T^\ell\}$ & $\bla$ & $\{-A, B\}$& $\{\alpha_\text{exp}, \beta_\text{exp}\}$ & $\E\T$ & $\E_i\T^\ell$ \\ \hline
     2-a & $0.01$  & $\substack{383\\400}$ & $\{{\bf 0}\}$& $\{{\bf 0}\}$ & $\substack{-4.06\\ 4.01}$  & $\substack{0.010\\0.010}$ & $17.32$ &  $\substack{\{12.05, 4.73, {\bf 0}\} \\\{6.15, 11.70, {\bf 0}\}}$   \\ \hline
    2-b & $0.01$  & $\substack{443\\450}$ & $\{7, 7, {\bf 0}\}$ & $\{0.1438, 0.137, {\bf 0}\}$ & $\substack{-4.02\\ 4.11}$  & $\substack{0.010\\ 0.010}$ & $17.59$ &  $\substack{\{9.31, 4.49, 2.97, 0.64\} \\\{4.95, 9.33, 0.78, 2.71\}}$   \\ \hline
  \end{tabular}
\end{table*}
\ignore{
\begin{table*}
\centering
\caption{{\color{red} This is table is for online experiment recording Fig. 3 (Unconstrained)}}
  \begin{tabular}{| c | c | c | c | c | c | c |}
    \hline
 Fig. & $\alpha=\beta$ & $\{\mu_0, \mu_1\}$ & $\{-A, B\}$& $\{\alpha_\text{exp}, \beta_\text{exp}\}$ & $\E\T$ & $\E_i\T^\ell$ \\ \hline
     3 & $0.1$  & $\substack{60\\60}$ & $\substack{-1.86\\ 1.87}$  & $\substack{0.100\\ 0.100}$ & $8.23$ &  $\substack{\{4.45, 3.785, {\bf 0}\}\\\{4.35, 3.70, {\bf 0}\} \\\{4.55, 3.87, {\bf 0}\}}$   \\ \hline
     3 & $0.08$  & $\substack{72\\72}$  & $\substack{-2.07\\ 2.07}$  & $\substack{0.079\\ 0.080}$ & $9.32$ &  $\substack{\{4.99, 4.33, {\bf 0}\}\\\{5.28, 3.92, {\bf 0}\} \\\{4.70, 4.74, {\bf 0}\}}$   \\ \hline
       3 & $0.06$  & $\substack{88\\88}$  & $\substack{-2.33\\ 2.31}$  & $\substack{0.060\\ 0.060}$ & $10.51$ &  $\substack{\{5.56, 4.955, {\bf 0}\}\\\{6.24, 4.2, {\bf 0}\} \\\{4.88, 5.71, {\bf 0}\}}$   \\ \hline
   
     3 & $0.04$  & $\substack{120\\125}$  & $\substack{-2.68\\ 2.59}$  & $\substack{0.039\\ 0.040}$ & $12.07$ &  $\substack{\{6.44, 5.63, {\bf 0}\}\\\{7.71, 4.22, {\bf 0}\} \\\{5.17, 7.04, {\bf 0}\}}$   \\ \hline

    3 & $0.02$  & $\substack{210\\220}$  & $\substack{-3.37\\ 3.36}$  & $\substack{0.02\\ 0.02}$ & $15.02$ &  $\substack{\{7.855, 7.16, {\bf 0}\}\\\{10.08, 4.86, {\bf 0}\} \\\{5.63, 9.46, {\bf 0}\}}$   \\ \hline
      3 & $0.01$  & $\substack{383\\400}$  & $\substack{-4.06\\ 4.01}$  & $\substack{0.010\\0.010}$ & $17.32$ &  $\substack{\{9.10, 8.215, {\bf 0}\}\\\{12.05, 4.73, {\bf 0}\} \\\{6.15, 11.70, {\bf 0}\}}$   \\ \hline
           
    3 & $0.008$  & $\substack{488\\520}$  & $\substack{-4.28\\ 4.33}$  & $\substack{0.079\\ 0.081}$ & $18.53$ &  $\substack{\{13.48, 5.03, {\bf 0}\} \\\{6.16, 12.39, {\bf 0}\}}$   \\ \hline
  3 & $0.006$  & $\substack{640\\688}$ & $\substack{-4.56\\ 4.55}$  & $\substack{0.006\\ 0.006}$ & $19.15$ &  $\substack{\{14.01, 4.92, {\bf 0}\} \\\{5.93, 13.43, {\bf 0}\}}$   \\ \hline
3 & $0.004$  & $\substack{915\\1010}$  & $\substack{-4.92\\ 5.05}$  & $\substack{0.0039\\ 0.0041}$ & $20.75$ &  $\substack{\{15.59, 4.99, {\bf 0}\} \\\{6.35, 14.57, {\bf 0}\}}$   \\ \hline
3 & $0.002$  & $\substack{1845\\2220}$  & $\substack{-5.62\\ 5.94}$  & $\substack{0.0019\\ 0.0021}$ & $23.33$ &  $\substack{\{18.11, 4.73, {\bf 0}\} \\\{6.78, 17.04, {\bf 0}\}}$   \\ \hline
  \end{tabular}
\end{table*}
}
\begin{table*}
\centering
\caption{{\color{red} This is table is for online experiment recording (constrained $N_1=6, N_2=9$)}}
  \begin{tabular}{| c | c | c | c | c | c | c | c |}
    \hline
 Fig. & $\alpha=\beta$ & $\{\mu_0, \mu_1\}$  & $\bla$ & $\{-A, B\}$& $\{\alpha_\text{exp}, \beta_\text{exp}\}$ & $\E\T$ & $\E_i\T^\ell$ \\ \hline
3 & $0.1$  & $\substack{60\\60}$ & $\{{\bf 0}\}$ & $\substack{-1.86\\ 1.87}$  & $\substack{0.100\\ 0.100}$ & $8.23$ &  $\substack{\{4.45, 3.785, {\bf 0}\}\\\{4.35, 3.70, {\bf 0}\} \\\{4.55, 3.87, {\bf 0}\}}$   \\ \hline
        3 & $0.08$  & $\substack{72\\72}$ & $\{{\bf 0}\}$ & $\substack{-2.07\\ 2.07}$  & $\substack{0.079\\ 0.080}$ & $9.32$ &  $\substack{\{4.99, 4.33, {\bf 0}\}\\\{5.28, 3.92, {\bf 0}\} \\\{4.70, 4.74, {\bf 0}\}}$  \\ \hline
     3 & $0.06$  & $\substack{88\\88}$ & $\{{\bf 0}\}$ & $\substack{-2.33\\ 2.31}$  & $\substack{0.060\\ 0.060}$ & $10.51$ &  $\substack{\{5.56, 4.95, {\bf 0}\}\\\{6.24, 4.2, {\bf 0}\} \\\{4.88, 5.71, {\bf 0}\}}$   \\ \hline

3 & $0.04$  & $\substack{120\\123}$ & $\{0.005, {\bf 0}\}$ & $\substack{-2.68\\ 2.59}$  & $\substack{0.039\\ 0.040}$ & $12.26$ &  $\substack{\{5.82, 6.44, {\bf 0}\}\\\{7.15, 5.16, {\bf 0}\} \\\{4.50, 7.73, {\bf 0}\}}$   \\ \hline

 3 & $0.02$  & $\substack{210\\230}$ & $\{0.085, {\bf 0}\}$ & $\substack{-3.39\\ 3.30}$  & $\substack{0.02\\ 0.02}$ & $14.86$ &  $\substack{\{6.01,  8.83, 0.01, 0\},\\ \{8.565, 6.976, 0.0006, 0\}\\\{3.465, 10.687, 0.0147, 0\}}$   \\ \hline

     3 & $0.01$  & $\substack{440\\480}$ & $\{0.15, 0.1115, {\bf 0}\}$ & $\substack{-4.09\\ 4.11}$  & $\substack{0.010\\0.010}$ & $17.93$ &  $\substack{\{5.97, 8.93, 2.55, 0.484\}\\\{7.99, 5.80, 4.23, 0.29\} \\\{3.94, 12.07,     0.874, 0.68\}}$   \\ \hline
    3 & $0.008$  & $\substack{546\\598}$ & $\{0.153, 0.1205, {\bf 0}\}$ & $\substack{-4.31\\ 4.35}$  & $\substack{0.080\\ 0.080}$ & $18.96$ &  $\substack{\{6.096, 8.976, 3.08, 0.80\}\\\{8.176, 5.59, 5.08, 0.42\} \\\{4.016, 12.362, 1.087, 1.185\}}$   \\ \hline

    3 & $0.006$  & $\substack{700\\800}$ & $\{0.1562,0.127, {\bf 0}\}$ & $\substack{-4.60\\ 4.63}$  & $\substack{0.0062\\ 0.0061}$ & $19.77$ &  $\substack{\{6.00, 8.94, 3.75,1.09\}\\\{8.30, 5.34, 6.02, 0.42\} \\\{3.69, 12.53,  1.47,     1.76\}}$   \\ \hline
    
        3 & $0.004$  & $\substack{1050\\1220}$ & $\{0.16125, 0.1365, {\bf 0}\}$ & $\substack{-4.99\\ 5.11}$  & $\substack{0.0039\\ 0.0039}$ & $21.43$ &  $\substack{\{5.95, 8.98, 4.85, 1.65\}\\\{8.30, 5.34, 6.02, 0.42\} \\\{3.69, 12.53,  1.47,     1.76\}}$   \\ \hline

3 & $0.002$  & $\substack{2035\\2695}$ & $\{0.17158, 0.1445, {\bf 0}\}$ & $\substack{-5.69\\ 5.94}$  & $\substack{0.0019\\ 0.0020}$ & $24.265$ &  $\substack{\{6.03, 8.95, 6.95, 2.33\}\\\{8.68, 4.78, 10.32    0.22\} \\\{3.39, 13.11,     3.58    4.45\}}$   \\ \hline

  \end{tabular}
\end{table*}
\ignore{
\begin{table*}
\centering
\caption{{\color{red} This is table is for offline experiment recording (Unconstrained)}}
  \begin{tabular}{ | c | c | c | c | c | c | c |}
    \hline
    $\alpha\, (\text{or}\,\beta)$ & $\{-A, B\}$& $\{\alpha_\text{exp}, \beta_\text{exp}\}$ & ${\bf p}$ & $\E\T$ & $\E\T^\ell (\E_0\T^\ell)$ \\ \hline
    $0.1$  & $\{-1.8, 1.8\}$ & $\{0.10, 0.10\}$ & $\{0.5, 0.5, {\bf 0}\}$ & $8.52$ & $\{4.256, 4.259, 0, 0\}$  \\ \hline
        $0.08$ &  $\{-2, 2\}$ & $\{0.080, 0.080\}$&$\{0.5, 0.5, {\bf 0}\}$ &$9.47$& $\{4.736, 4.736, {\bf 0}\}$ \\ \hline
    $0.06$ &  $\{-2.32, 2.31\}$ & $\{0.060, 0.060\}$&$\{0.5, 0.5, {\bf 0}\}$ &$11.00$& $\{5.5, 5.5, {\bf 0}\}$ \\ \hline
    $0.04$ &  $\{-2.7, 2.7\}$ & $\{0.040, 0.040\}$&$\{0.5, 0.5, {\bf 0}\}$ &$12.88$& $\{6.44, 6.44, {\bf 0}\}$ \\ \hline
    $0.02$ &  $\{-3.4, 3.4\}$ & $\{0.020, 0.020\}$&$\{0.5, 0.5, {\bf 0}\}$ &$16.00$& $\{8.00, 8.00, {\bf 0}\}$ \\ \hline
       $0.01$ &  $\{-4.1, 4.08\}$ & $\{0.01, 0.01\}$&$\{0.5, 0.5, {\bf 0}\}$ &$18.982$& $\{9.4833, 9.4991, {\bf 0}\}$ \\ \hline
     $0.008$ &  $\{-4.3, 4.3\}$ & $\{0.0082, 0.0080\}$&$\{0.5, 0.5, {\bf 0}\}$ &$19.91$& $\{9.95, 9.955, {\bf 0}\}$ \\ \hline
    $0.006$ &  $\{-4.6, 4.6\}$ & $\{0.0056, 0.0060\}$&$\{0.5, 0.5, {\bf 0}\}$ &$21.22$& $\{10.62, 10.61, {\bf 0}\}$ \\ \hline
    $0.004$ & $\{-5.05, 5.05\}$& $\{0.0039, 0.0036\}$&$\{0.5, 0.5, {\bf 0}\}$& $23.06$&$\{11.52, 11.53, {\bf 0}\}$ \\ \hline
    $0.002$ & $\{-5.83, 5.83\}$& $\{0.0019, 0.0020\}$&$\{0.5, 0.5, {\bf 0}\}$& $26.22$&$\{13.10, 13.12, {\bf 0}\}$ \\ \hline
\end{tabular}
\end{table*}
}
\begin{table*}
\centering
\caption{{\color{red} This is table is for offline experiment recording (Constrained: $N_1=6, N_2=9$)}}
  \begin{tabular}{ | c | c | c | c | c | c | c |}
    \hline
    $\alpha\, (\text{or}\,\beta)$ & $\{-A, B\}$& $\{\alpha_\text{exp}, \beta_\text{exp}\}$ & ${\bf p}$ & $\E\T$ & $\E\T^\ell (\E_0\T^\ell)$ \\ \hline
    $0.1$  & $\{-1.8, 1.8\}$ & $\{0.10, 0.10\}$ & $\{0.5, 0.5, {\bf 0}\}$ & $8.52$ & $\{4.256, 4.259, 0, 0\}$  \\ \hline
        $0.08$ &  $\{-2, 2\}$ & $\{0.080, 0.080\}$&$\{0.5, 0.5, {\bf 0}\}$ &$9.47$& $\{4.736, 4.736, {\bf 0}\}$ \\ \hline
    $0.06$ &  $\{-2.32, 2.31\}$ & $\{0.060, 0.060\}$&$\{0.5, 0.5, {\bf 0}\}$ &$11.00$& $\{5.5, 5.5, {\bf 0}\}$ \\ \hline
    $0.04$ &  $\{-2.7, 2.7\}$ & $\{0.041, 0.041\}$&$\{0.475, 0.5, 0.015, 0\}$ &$12.88$& $\{5.99, 6.44, 0.453, 0\}$ \\ \hline
    $0.02$ &  $\{-3.4, 3.4\}$ & $\{0.020, 0.020\}$&$\{0.372, 0.5, 0.128, 0\}$ &$16.23$& $\{6.036, 8.1253, 2.073, 0\}$ \\ \hline
    $0.01$ &  $\{-4.1, 4.08\}$ & $\{0.01, 0.01\}$&$\{0.31, 0.465, 0.16, 0.065\}$ &$19.4280$& $\{6.0259, 9.0397, 3.0985, 1.2639\}$ \\ \hline
     $0.008$ &  $\{-4.3, 4.3\}$ & $\{0.0081, 0.0081\}$&$\{0.29, 0.435, 0.15, 0.125\}$ &$20.4767$& $\{5.994, 9.00, 4.51, 0.971\}$ \\ \hline
    $0.006$ &  $\{-4.6, 4.6\}$ & $\{0.006, 0.0060\}$&$\{0.275, 0.4125, 0.18, 0.1325\}$ &$21.88$& $\{6.0176, 9.0265, 3.941,2.9\}$ \\ \hline
    $0.004$ & $\{-5.05, 5.05\}$& $\{0.0040, 0.0041\}$&$\{0.25, 0.375, 0.18, 0.195{\bf 0}\}$& $23.9722$&$\{5.992, 8.994, 4.313, 4.673\}$ \\ \hline
    $0.002$ & $\{-5.83, 5.83\}$& $\{0.0020, 0.0020\}$&$\{0.218, 0.327, 0.22, 0.235\}$& $27.537$&$\{6.01, 9.002, 6.0494, 6.4762\}$ \\ \hline
\end{tabular}
\end{table*}

\begin{table*}
\centering
\caption{{\color{red} Infinite-Horizon: This is table is for online experiment recording (constrained $N_1=6, N_2=9$)}}
  \begin{tabular}{| c | c | c | c | c | c | c | c |}
    \hline
 $N$ & $\alpha=\beta$ & $\{\mu_0, \mu_1\}$  & $\bla$ & $\{-A, B\}$& $\{\alpha_\text{exp}, \beta_\text{exp}\}$ & $\E\T$ & $\E_i\T^\ell$ \\ \hline
150 & $0.1$  & $\substack{60\\60}$ & $\{{\bf 0}\}$ & $\substack{-1.86\\ 1.87}$  & $\substack{0.100\\ 0.100}$ & $8.23$ &  $\substack{\{4.45, 3.785, {\bf 0}\}\\\{4.35, 3.70, {\bf 0}\} \\\{4.55, 3.87, {\bf 0}\}}$   \\ \hline
  150 & $0.08$  & $\substack{72\\72}$ & $\{{\bf 0}\}$ & $\substack{-2.07\\ 2.07}$  & $\substack{0.079\\ 0.080}$ & $9.32$ &  $\substack{\{4.99, 4.33, {\bf 0}\}\\\{5.28, 3.92, {\bf 0}\} \\\{4.70, 4.74, {\bf 0}\}}$  \\ \hline
 150 & $0.06$  & $\substack{88\\88}$ & $\{{\bf 0}\}$ & $\substack{-2.31\\ 2.33}$  & $\substack{0.060\\ 0.061}$ & $10.42$ &  $\substack{\{5.5, 4.92, {\bf 0}\}\\\{6.15, 4.05, {\bf 0}\} \\\{4.855, 5.78, {\bf 0}\}}$   \\ \hline

150 & $0.04$  & $\substack{120\\123}$ & $\{0.005, {\bf 0}\}$ & $\substack{-2.70\\ 2.69}$  & $\substack{0.040\\ 0.040}$ & $12.25$ &  $\substack{\{5.88, 6.375, {\bf 0}\}\\\{7.23, 5.054, {\bf 0}\} \\\{4.533, 7.70, {\bf 0}\}}$   \\ \hline

 200 & $0.02$  & $\substack{210\\230}$ & $\{0.085, {\bf 0}\}$ & $\substack{-3.39\\ 3.30}$  & $\substack{0.02\\ 0.02}$ & $14.82$ &  $\substack{\{5.975, 8.834, 0.0141, 0\},\\ \{8.50, 7.00, 0.0011, 0\}\\\{3.451, 10.670, 0.0271, 0\}}$   \\ \hline

    200 & $0.01$  & $\substack{420\\460}$ & $\{0.1470, 0.105, {\bf 0}\}$ & $\substack{-4.04\\ 4.06}$  & $\substack{0.010\\0.010}$ & $17.62$ &  $\substack{\{5.973, 9.026, 2.320, 0.302\}\\\{8.1758, 6.0010, 3.840, 0.166\} \\\{3.770, 12.043, 0.800, 0.438\}}$   \\ \hline
    200 & $0.008$  & $\substack{520\\570}$ & $\{0.1512, 0.1160, {\bf 0}\}$ & $\substack{-4.2637\\ 4.3}$  & $\substack{0.080\\ 0.081}$ & $18.57$ &  $\substack{\{5.93, 8.96, 3.026, 0.65\}\\\{7.941, 5.659, 5.036, 0.3146\} \\\{3.911, 12.293, 1.016, 0.983\}}$   \\ \hline

    200 & $0.006$  & $\substack{690\\780}$ & $\{0.1535,0.1265, {\bf 0}\}$ & $\substack{-4.58\\ 4.62}$  & $\substack{0.0062\\ 0.0061}$ & $20.066$ &  $\substack{\{6.0321, 9.0176, 3.8684, 1.1476\}\\\{8.2862, 5.5553, 6.2202, 0.4541\} \\\{3.778,    12.480, 1.5165, 1.841\}}$   \\ \hline
    
        200 & $0.004$  & $\substack{1000\\1180}$ & $\{0.1562, 0.1346, {\bf 0}\}$ & $\substack{-4.97\\ 5.05}$  & $\substack{0.0039\\ 0.0039}$ & $21.36$ &  $\substack{\{6.06, 9.026, 4.57, 1.65\}\\\{8.66, 5.402, 7.2165, 0.399\} \\\{3.562, 12.651, 1.926, 2.905\}}$   \\ \hline

200 & $0.002$  & $\substack{1860\\2580}$ & $\{0.1615, 0.1373, {\bf 0}\}$ & $\substack{-5.65\\ 5.81}$  & $\substack{0.0019\\ 0.0019}$ & $24.217$ &  $\substack{\{6.011, 9.02, 6.913, 2.264\}\\\{9.1554, 4.7665, 10.178, 0.253\} \\\{2.867,   13.2926, 3.649, 4.274\}}$   \\ \hline

  \end{tabular}
\end{table*}
\begin{table*}
\centering
\caption{{\color{red} Infinite-Horizon: This is table is for offline experiment recording (Constrained: $N_1=6, N_2=9$)}}
  \begin{tabular}{ | c | c | c | c | c | c | c |}
    \hline
    $\alpha\, (\text{or}\,\beta)$ & $\{-A, B\}$& $\{\alpha_\text{exp}, \beta_\text{exp}\}$ & ${\bf p}$ & $\E\T$ & $\E\T^\ell (\E_0\T^\ell)$ \\ \hline
    $0.1$  & $\{-1.8, 1.8\}$ & $\{0.10, 0.10\}$ & $\{0.5, 0.5, {\bf 0}\}$ & $8.52$ & $\{4.256, 4.259, 0, 0\}$  \\ \hline
        $0.08$ &  $\{-2, 2\}$ & $\{0.080, 0.080\}$&$\{0.5, 0.5, {\bf 0}\}$ &$9.47$& $\{4.736, 4.736, {\bf 0}\}$ \\ \hline
   $0.06$ &  $\{-2.32, 2.31\}$ & $\{0.060, 0.060\}$&$\{0.5, 0.5, {\bf 0}\}$ &$11.00$& $\{5.5, 5.5, {\bf 0}\}$ \\ \hline
   $0.04$ &  $\{-2.7, 2.7\}$ & $\{0.041, 0.041\}$&$\{0.472, 0.5, 0.028, 0\}$ &$12.838$& $\{6.0575, 6.425, 0.3557, 0\}$ \\ \hline
  $0.02$ &  $\{-3.4, 3.35\}$ & $\{0.020, 0.020\}$&$\{0.373, 0.6, 0.027, 0\}$ &$15.98$& $\{5.960, 9.5873, 0.4333, 0\}$ \\ \hline
   $0.01$ &  $\{-4.1, 4.1\}$ & $\{0.01, 0.01\}$&$\{0.31, 0.465, 0.16, 0.065\}$ &$19.430$& $\{6.048, 9.080, 3.123, 1.270\}$ \\ \hline
    $0.008$ &  $\{-4.3, 4.3\}$ & $\{0.0081, 0.0081\}$&$\{0.29, 0.435, 0.15, 0.125\}$ &$20.477$& $\{5.994, 9.00, 4.51, 0.971\}$ \\ \hline
    $0.006$ &  $\{-4.6, 4.6\}$ & $\{0.0061, 0.0060\}$&$\{0.274, 0.411, 0.250, 0.065\}$ &$21.74$& $\{5.963, 8.9243, 5.4415, 1.4113\}$ \\ \hline
    $0.004$ & $\{-5.00, 5.00\}$& $\{0.0040, 0.0041\}$&$\{0.253, 0.3795, 0.210,    0.1575\}$& $23.71$&$\{5.992, 9.002, 4.983, 3.734\}$ \\ \hline
    $0.002$ & $\{-5.72, 5.72\}$& $\{0.0019, 0.002\}$&$\{0.223, 0.3345, 0.330, 0.1125\}$& $27.06$&$\{6.0317, 9.0467, 8.934, 3.0477\}$ \\ \hline
\end{tabular}
\end{table*}}
\ignore{
\section{Optimal Adaptive Scheme with Usage Constraints---Scenario II}
{\color{red} Start with this problem. Prove the structure with a general theorem. Interpret the theorem in a more general application background (general cost). }
\begin{align}\label{P2}
\begin{array}{cl}
\min_{\{\bs_1^\infty, D, \T\}} & \E^{\bs_1^\infty}\lb c\T +\mathbbm{1}_{\{D\neq H\}}\rb\\
\text{subject to} & \E^{\bs_1^\infty}\lb \sum_{t=1}^\T \bs_t(\ell)\rb\le \xi_\ell \,\E^{\bs_1^\infty}\T, \quad \ell=1, \ldots, L.\tag{P2}
\end{array}
\end{align}

{\color{red} Shall we consider a general theorem here, or just consider a specific problem. At least, the SPRT structure remains the same for general problem.
The optimal structure holds for any selection strategy. We then can discuss the particular case where one sensor is selected at a time, that could correspond to one mode is switched on at each time.  }

Note that the utilization constraints are on average sense. For instance, the sensor system can be resused and the average battery life imposes the constraint on each sensor. {\color{red} The constraint on the utilization percentage. } Here $\sum_{\ell=1}^L \xi_\ell\ge 1$ is necessary because we assume that at each time at least one sensor must be selected, i.e., $\sum_{\ell=1}^L\bs_t(\ell)\ge 1$ for $t=1, \ldots, \T$, thus
\begin{align}\label{eq1}
\sum_{\ell=1}^L\xi_\ell\,\E^{\bs_1^\infty}\T\ge \E^{\bs_1^\infty}\lb\sum_{t=1}^\T\sum_{\ell}^L\bs_t(\ell)\rb\ge\E^{\bs_1^\infty}\lb\T\rb,
\end{align}
where the second equality holds true when $\sum_{\ell}^L\bs_t(\ell)=1$, only one sensor is used at each time. If $\sum_{\ell=1}^L\xi_\ell=1$, then only one sensor is active at each time, and all constraints are satisfied with equality. This is true because $$\E^{\bs_1^\infty}\T\ge \E^{\bs_1^\infty}\lb\sum_{t=1}^\T\sum_{\ell=1}^L\bs_t(\ell)\rb\ge \E^{\bs_1^\infty}\lb\T\rb\to \E^{\bs_1^\infty}\lb\sum_{t=1}^\T\sum_{\ell=1}^L\bs_t(\ell)\rb= \E^{\bs_1^\infty}\lb\T\rb$$
i.e., the second inequality in \eqref{eq1} holds as equality, thus $\sum_{\ell}^L\bs_t(\ell)=1$ (one sensor is active at each time). If there exists a sensor $\ell$ such that $\E^{\bs_1^\infty}\sum_{t=1}^\T\bs_t(\ell)<\xi_\ell\, \E^{\bs_1^\infty}\T$, then
$$\sum_{\ell=1}^L\E^{\bs_1^\infty}\sum_{t=1}^\T\bs_t(\ell)<\E^{\bs_1^\infty}\T,$$ and contradiction occurs. By introducing the multipliers, \eqref{P1} becomes
\begin{align}
\min_{\{\bs_1^\infty, D, \T\}}\quad \E\lb c\T +\mathbbm{1}_{\{D\neq H\}}\rb+ c\sum_{\ell=1}^L\lambda_\ell\,\E\lb \sum_{t=1}^\T \lb\bs_t(\ell)-\xi_\ell\rb\rb, \quad \lambda_\ell\ge 0.
\end{align}
Here the factor $c$ in the second term is introduced for presentation conciseness in the forthcoming derivations. We can further write the objective function as
\begin{align}
 &c\,\E\T+\E\lb \mathbbm{1}_{\{D\neq H\}}\rb+c\,\E\lb\sum_{t=1}^\T\sum_{\ell=1}^L\lambda_\ell\,(\bs_t(\ell)-\xi_\ell)\rb\nonumber\\=&\,\E\lb c \sum_{t=1}^\T \lb 1+\sum_{\ell=1}^L\lambda_\ell\,\bs_t(\ell)-\sum_{\ell=1}^L\lambda_\ell\xi_\ell\rb+\mathbbm{1}_{\{D\neq H\}}\rb
\end{align}

{\color{red} Any insight compared to (P1)?}
}
\bibliographystyle{IEEEtran}
\bibliography{IEEEabrv,references}
\end{document}